\documentclass[12pt]{article}
\usepackage{authblk}
\usepackage[top=2cm, bottom=2cm, left=2cm, right=2cm]{geometry}
\usepackage{amsmath,amssymb,amsthm}
\usepackage{hyperref}
\usepackage[dvipsnames]{xcolor}
\usepackage{graphicx}
\usepackage{caption}
\usepackage{subcaption}
\usepackage{lineno}
\usepackage[title]{appendix}
\usepackage{soul}
\usepackage{multirow,multicol,booktabs}
\usepackage{natbib}

\title{Comparing regional and provincial-wide COVID-19 models with physical distancing in British Columbia}

\author[1]{Geoffrey McGregor}
\author[2]{Jennifer Tippett}
\author[1]{Andy T.S. Wan\footnote{Author for correspondence. Email: andy.wan@unbc.ca}}
\author[3]{Mengxiao Wang}
\author[3]{Samuel W.K. Wong}

\affil[1]{\footnotesize{Department of Mathematics and Statistics, University of Northern British Columbia, Prince George, Canada}}
\affil[2]{\footnotesize{School of Health Sciences, University of Northern British Columbia, Prince George, Canada}}
\affil[3]{\footnotesize{Department of Statistics and Actuarial Science, University of Waterloo, Waterloo, Canada}}
\date{November 7, 2021}                     
\begin{document}
  \maketitle
\begin{abstract}
We study the effects of physical distancing measures for the spread of COVID-19 in regional areas within British Columbia, using the reported cases of the five provincial Health Authorities. Building on the Bayesian epidemiological model of \cite{AndersonAE20}, we propose a hierarchical regional Bayesian model with time-varying regional parameters between March to December of 2020. In the absence of COVID-19 variants and vaccinations during this period, we examine the regionalized basic reproduction number, modelled prevalence, relative reduction in contact due to physical distancing, and proportion of anticipated cases that have been tested and reported. We observe significant differences between the regional and provincial-wide models and demonstrate the hierarchical regional model can better estimate regional prevalence, especially in rural regions. These results indicate that it can be useful to apply similar regional models to other parts of Canada or other countries.
\end{abstract}

\section{Introduction}

The coronavirus disease 2019 (COVID-19) is caused by the SARS-CoV-2 virus, and was declared by the World Health Organization to be a global pandemic on March 11, 2020. The virus and its associated illness have spread from their origin in Wuhan, China to nearly every corner of the world. The first reported case of COVID-19 arrived in the province of British Columbia (BC) in Canada on January 28, 2020, and on the one-year anniversary of the pandemic, BC has had 86,219 positive cases and 1,397 deaths, as reported by the \cite{GC21a}. 

Prior to the development of vaccines and also with on-going COVID-19 immunizations, many countries around the world and their associated provinces, states, or sub-regions, including BC, have been relying on non-medical interventions to control the transmission of the virus. The application of these interventions — including  mask usage, physical distancing, contact tracing, hand washing, improved ventilation recommendations, travel restrictions, and restriction on social gatherings — can be either fixed or dynamic and is intended to prevent overwhelming the health care system. Fixed interventions are in place for a set duration of time and dynamic interventions are cycled on and off in response to the demand on the health care system. Periods of tightening and relaxation of COVID-19 restrictions can have a dramatic effect on the number of new COVID-19 case incidences over time and the overall duration of the pandemic as discussed in \cite{ShaferNescaBalshaw21}. Thus some countries, including New Zealand, have preferred a ``COVID zero'' approach where these non-medical interventions are strictly enforced with the goal of eradication of COVID-19 within a specific geographic region, as recommended by \cite{Bakerm4907}.
In contrast, BC has applied dynamic non-medical interventions to control the strain on the health care system for the majority of 2020.  

Urban and rural areas, including those within BC, are heterogeneous in terms of population density, cultural profile, social determinants of health, and attitudes towards COVID-19 restrictions. As defined by \cite{HCSocial}, social determinants of health refer to social and economic factors affecting the health of groups of individuals.  
Specifically, there is an inherent lack of availability of health services in rural areas and significant geographical barriers to accessing health services. On average, rural residents have to travel four times farther than those in urban regions to visit a physician, as reported in \cite{GarasiaDobb19}. Moreover, non-medical interventions rely on the collective goodwill of the people to comply with public health orders and the perceived threat of COVID-19. These behaviours and perceptions are known to vary between urban and rural regions. Specifically, rural residents have been found to be less likely to adapt their behaviour in response to COVID-19 and to have more negative perceptions of the effectiveness of the non-medical interventions, as discussed in \cite{ChenChen20}. While these significant differences between urban and rural areas persist, BC has seen predominantly blanketed restrictions across the province for the majority of 2020, based on COVID-19 projections using data sourced from urban centres. Thus, macro-level modelling at the provincial or territorial level may be ignoring the nuances of rural living and limiting the specificity of these models to non-urban areas. 
One metric utilized for quantifying these regional differences is the basic reproduction number, $R_0$, defined as the expected number of new cases a single infectious individual will generate. As discussed in \cite{Delamater19}, a region's estimated $R_0$ value will vary depending on the population density and social behaviours specific to that region. For example, in a COVID-19 modelling study of \cite{TurkAE20}, $R_0$ was estimated to be lower for a central urban area than for the entire state of North Carolina.  

In BC, a compartmental mathematical modelling study by \cite{AndersonAE20} played an instrumental role in guiding the reopening plans devised by the provincial government after it shut down all but essential services on March 17, 2020. This model was used to assess the spread of COVID-19 in BC and to provide projections based on various levels of physical distancing. It was constructed as a modified SEIR model, which differs from typical SEIR models in that each model compartment was replicated, in order to create a series of two parallel compartments for those who did and did not engage in physical distancing. While the model of \cite{AndersonAE20} was vital in BC’s response to COVID-19 at the beginning of the pandemic, it did not account for the inherent and aforementioned differences between urban and rural regions. Furthermore, it contained parameters related to physically distancing that were derived from urban data sources, such as the rate at which individuals move to and from being physically distanced.

In this work, there are two main objectives. First, we adapt the compartmental model of \cite{AndersonAE20} to a regional model. Second, we study the differences between the proposed regional model and the provincial-wide model during March to December of 2020, prior to the vaccination phase beginning in January of 2021. Specifically, we propose a hierarchical Bayesian epidemiological model of COVID-19 for each of BC’s five regional health authorities: Vancouver Coastal Health, Fraser Health, Interior Health, Island Health, and Northern Health. We refer the five regional health authorities respectively as Coastal, Fraser, Interior, Island and Northern and often refer to them as regions. The former two regions contain the most populated urban areas in BC, including Greater Vancouver and the Fraser Valley, whereas the latter three regions are primarily rural and less populated. Since previous research has highlighted the inverse relationship between rurality and the likelihood of modified behaviours in response to the pandemic, certain model parameters are regionalized to these five regions of BC. Specifically, we chose the regional parameters to be the \emph{relative reduction in contacts due to physical distancing} and the \emph{proportion of anticipated COVID-19 cases that have been tested and reported}, leading to a Bayesian hierarchical model.  

We discuss briefly the assumptions of our hierarchical regional model and limitations to our approach, for which the subsequent section will provide more details. While grouping the provincial case counts into regions would allow us to inspect case data more precisely within each region,  looking solely at the regional data can be misleading for predicting future regional case counts. One main reason is that these regions are not isolated, since there can be migration between regions of British Columbia, even during periods of intraprovincial travel restrictions. 
Therefore, if one region has a large spike in cases, then it is possible that neighbouring regions will also see an increase in cases over time. The impacts of migration on COVID-19 have been recently studied by \cite{hu2021population, sirkeci2020coronavirus, zhan2020general}. Overall, reducing regional migration is an important aspect of controlling the spread of COVID-19. Despite migration playing a significant role in the spread of COVID-19 throughout regions, an epidemiological and statistical model accounting for migration requires access to migration data, which were not publicly available for the period of our study. Thus, we do not inherently take into account of migration between regions and instead use the observed case counts to infer estimates about our model parameters. 
As shown later in the Results and Discussion section, we observed that there were periods when the provincial trend was opposite to that of specific regional trends. For example, the province as a whole may be experiencing a rapid decrease in daily cases, while certain rural regions are experiencing growth in cases. In these situations, the insight gained from our regional modelling may help guide region specific mitigation measures. Conversely, if the opposite scenario occurs, where the provincial case numbers are on the rise while certain regions' case numbers remain low, then restricting migration may help to prevent the spread of new cases into these regions.

This paper is organized as follows. In Section \ref{Modelling}, we extend the Bayesian epidemiological model introduced by \cite{AndersonAE20} to include the five health regions of BC. Specifically, we introduce a regional epidemiological model in Section \ref{Epi Model} and discuss the regional time-varying specific parameters, such as regional relative contact reduction due to physical distancing. In Section \ref{Stat Model}, we detail the hierarchical structure of our Bayesian model. In particular, we discuss the regional proportion of anticipated cases that have been tested and reported due to improvement in testing and provide justifications on the choice of priors. In Section \ref{sec:modelInput}, we summarize specific model parameters for both the hierarchical regional model and the provincial-wide model used for comparison. In Section \ref{sec:simstudy}, we assess the hierarchical model's ability to estimate model parameters from a controlled simulation study. In Section \ref{sec:Just}, we provide justifications on our choice of hierarchical model versus a non-hierarchical one. Finally, in Section \ref{results}, we present our main results and compare the differences in prevalence and parameters between the provincial-wide and regional models. To close the paper, we summarize our work and discuss limitations of our regional model, along with future work.

\section{Modelling}
\label{Modelling}
In this section, we introduce an extension of the mathematical and statistical framework presented in \cite{AndersonAE20} for modelling the spread of COVID-19 throughout BC, Canada. We choose to build upon this model as it captures many of the key features required for modelling COVID-19. For example, the epidemiological model contains COVID-19 specific compartments, such as, exposed but not infectious, infectious but not symptomatic, and quarantining. In addition to these compartments, populations practising physical distancing are accounted for, with reductions in exposure being time-dependent based on government lockdown protocols.  The model also accounts for changes in test availability by allowing the proportion of infected individuals being tested to change over time while also accounting for potential delays between the onset of symptoms and getting a positive test. The resulting epidemiological model is then utilized within the statistical framework to compute likely values of the unknown model parameters. As in \cite{AndersonAE20},  we take a Bayesian approach to parameter estimation and sample values of parameters from their posterior distribution via Hamiltonian Monte Carlo (HMC).  Computations are carried out in R, using the Stan package, as detailed in \cite{carpenter2017stan}.  Further details on the statistical approach are in Section \ref{Stat Model}.

Before we proceed to the model specifics, we highlight our proposed extensions to the model presented in \cite{AndersonAE20}. First and foremost, we regionalize the model to describe the spread of COVID-19 in each of the five health regions of BC: Coastal, Fraser, Interior, Island, and Northern. This entails having a distinct set of differential equations for each of these regions, with certain parameters being region-specific and others shared provincial-wide. We also extend the modelling period to the end of December 2020, therefore we have included seven different physical distancing periods, based on government protocol, and four different testing periods accounting for the increasing availability of tests. 

We have restricted our modelling period from March to December 2020 for two main reasons. Beginning in January 2021, BC began rolling out vaccines, as reported by \cite{CBCNews21}. Accounting for vaccinated individuals would require alterations to the epidemiological model. While such extensions are well studied in the literature, it is not the main goal of our work. In addition to vaccinations, COVID-19 variants began spreading throughout BC and the rest of Canada starting in mid-December of 2020, as reported by \cite{CityNews21}. Moreover, certain variants are known to be far more infectious according to \cite{BCCDC21b}, and therefore, significant changes to the modelling framework would be required to study the impacts of the new variants. Given that our main objective is to study the impacts of regional versus provincial-wide modelling, we have focused our attention on data from the time period between March and December of 2020.

\subsection{Epidemiological Model}\label{Epi Model}
We now present a regionalized version of the epidemiological model of \cite{AndersonAE20} and discuss the assumptions of our modelling choices throughout. 

The epidemiological model being considered is an SEIQR type model, meaning we model individuals in health region $i$ which are susceptible ($S^i$), exposed, not symptomatic and not infectious ($E^i_1$), exposed, not symptomatic and infectious ($E^i_2$), symptomatic and infectious ($I^i$), quarantined ($Q^i$) and recovered ($R^i$).   In addition to these compartments, a secondary set of equations $S^i_d$, $E^i_{1d}$, $E^i_{2d}$, $I^i_{d}$, $Q^i_{d}$ and $R^i_d$, is included to model individuals in health region $i$ \emph{practising physical distancing}. The strictness of the physical distancing measures will vary in time due to policy changes throughout the different lockdown phases between March 1, 2020 and December 31, 2020. As discussed in more detail later in this section, the proposed epidemiological model has no population influx or out-flux, meaning \emph{births and deaths are not being accounted for}.  
Due to the relatively low death rate, and the relatively short time period being studied, we believe this is a reasonable modelling choice.

Within the $i^{th}$ health region with population $N^i$, the epidemiological model for individuals not practising physical distancing is given by the ordinary differential equations (ODEs):

\begin{align}
\begin{split}
\frac{dS^i}{dt}&=-\beta \left(I^i+E^i_2+f^i(t)(I^i_d+E^i_{2d})\right)\frac{S^i}{N^i}-u_dS^i+u_rS^i_d \label{Model} \\
\frac{d E^i_1}{dt}&=\beta \left(I^i+E^i_2+f^i(t)(I^i_d+E^i_{2d})\right)\frac{S^i}{N^i}-k_1E^i_1-u_dE^i_1+u_rE^i_{1d}  \\
\frac{d E^i_2}{dt}&=k_1E^i_1-k_2E^i_2-u_dE^i_2+u_rE^i_{2d} \\
\frac{d I^i}{dt}&=k_2E^i_2-qI^i-\frac{1}{D}I^i-u_dI^i+u_rI^i_{d} \\
\frac{d Q^i}{dt}&=qI^i-\frac{1}{D}Q^i-u_dQ^i+u_rQ^i_{d} \\
\frac{d R^i}{dt}&=\frac{1}{D}(I^i+Q^i)-u_dR^i+u_rR^i_{d}.
\end{split}
\end{align}
For ease of reference and clarity, we have summarized the provincial and regional parameters appearing in Equation (\ref{Model}) in Table \ref{table:constants}. The values for the fixed model parameters $k_1, k_2, D, q, u_r$ and $u_d$ were chosen to be the same as in \cite{AndersonAE20}.

\begin{table}[h!]
\centering
\begin{tabular}{|c|c|c|}
\hline
Model Parameter & Type & Value\\
\hline
\hline
$k_1$ & Provincial & 0.2 days$^{-1}$\\
\hline
$k_2$ & Provincial & 1 days$^{-1}$\\
\hline
$D$ & Provincial & 5 days\\
\hline
$q$ & Provincial & 0.05 \\
\hline
$u_r$ & Provincial & 0.1\\
\hline
$u_d$ & Provincial & 0.02\\
\hline
$f^i$ & Regional & Estimated \\
\hline
$\beta$ & Provincial & Estimated \\
\hline
\end{tabular}
\caption{Epidemiological model parameters}
\label{table:constants}
\end{table}

The parameters  $k_1$, $k_2$ and $\frac{1}{D}$ quantify the progression of the infected population through the compartments $E^i_1$ to $E^i_2$ and eventually to $I^i$ and $R^i$. Specifically, $D$ represents the mean duration of the infectious period, and the parameter $q$  represents the rate at which infectious individuals begin fully quarantining. We have chosen to fix $k_1, k_2$ and $D$ throughout the province, as they are intrinsic  to COVID-19 during the period of our study and we fix $q$ as we do not have access to region specific quarantining data.  As discussed in \cite{AndersonAE20}, approximately one fifth of all severe cases eliminated transmission by either fully quarantining, or ending up in the hospital (both considered to be within compartment $Q^i$). Therefore, equating $\frac{1}{5}=\frac{q}{q+1/D}$, yields the value $q=0.05$ as seen in Table \ref{table:constants}. The parameter $u_d$ governs the rate at which individuals switch their behaviour to begin physical distancing and $u_r$ governs the rate at which they return to normal behaviour\footnote{While ideally both $u_r$ and $u_d$ should be made region specific and evolve over time, however, we do not have access to such regional data required for such a modelling choice. As a result, we have kept $u_r$ and $u_d$ common throughout the province.}.  Later in this section, we also show how $u_r$ and $u_d$ describe the asymptotic proportion of the population who are practising physical distancing. Since the physical distancing mandates were provincial-wide during the time period under study, $u_d$ and $u_r$ are taken as provincial-wide fixed parameters. Furthermore, we interpret the parameters $f^i$ as \emph{the relative reduction in contact due to physical distancing in region $i$} and $\beta$ as \emph{the rate of transmission from the infectious population to the susceptible population}. We briefly discuss these interpretations next.
 
Firstly, physical distancing has not been the only tactic employed throughout British Columbia to mitigate the spread of COVID-19. Other mitigation measures, such as mask usage, contact tracing, hand washing, sanitization and improved ventilation recommendations, also played prominent roles in mitigating transmission of COVID-19 between March and December of 2020. Therefore, if a lower value of $f^i(t)$ is observed in a region, this could be interpreted as a combination of mitigation factors, instead of a relative reduction in contacts based solely on physical distancing. However, the lockdown phases under study were mostly associated with changes in physical distancing protocols within the province. The other aforementioned mitigation measures remained largely unchanged throughout this same time period. Thus, while a baseline portion of $f^i(t)$ can be attributed to other mitigation measures, the changes in observed case counts in the provincial and regional data are then largely due to the changes in physical distancing measures. Due to lack of data to account for the effects of other mitigation measures on $f^i(t)$, we therefore make the assumption that \emph{physical distancing is the largest contributor to mitigating transmission of COVID-19 during the time period under study}. Thus, for the rest of the paper, we interpret $f^i(t)$ mostly as the relative reduction in contacts due to physical distancing. Furthermore, this is also consistent with the interpretation of $f(t)$ as presented in the original model of \cite{AndersonAE20}.
 
Secondly, given the heterogeneity within the regions of British Columbia, it seems reasonable to have the rate of transmission parameter, $\beta$, be region specific. However, there is dependence between $\beta$ and the relative reduction in contacts, $f^i(t)$, leading to some over-parametrization in our regional model in the absence of additional constraints. For example, the same case count data can be modelled by different combinations of parameter values:  (a) A less contagious disease (smaller $\beta$) but with more contacts (larger $f^i(t)$), or (b) A more contagious disease and less contacts.  
 Given that we are interested in observing the changes in $f^i(t)$ over time, and comparing these values across regions, its dependence on $\beta$ complicates their interpretation. We therefore have chosen to estimate \emph{$\beta$ as a provincial-wide parameter and $f^i(t)$ as regional parameters.}  Prior distributions can be assigned to stabilize these parameters while providing sufficient flexibility for the data to inform their estimates; these are discussed in Section \ref{Stat Model}. 
  The justification of this modelling choice is further explored in more detail in Section \ref{sec:Just}. 

Having discussed these assumptions and limitations, we continue with introducing the remaining part of the model for the analogous compartments practising physical distancing. Specifically, within the $i^{th}$ health region, individuals practising physical distancing have their own set of ODEs, given by:

\begin{align}
\begin{split}
\frac{dS^i_d}{dt}&=-f^i(t)\beta \left(I^i+E^i_2+f^i(t)(I^i_d+E^i_{2d})\right)\frac{S^i_d}{N^i}+u_dS^i-u_rS^i_d\label{SDModel} \\
\frac{d E^i_{1d}}{dt}&=f^i(t)\beta \left(I^i+E^i_2+f^i(t)(I^i_d+E^i_{2d})\right)\frac{S^i_d}{N^i}-k_1E^i_{1d}+u_dE^i_1-u_rE^i_{1d}\\
\frac{d E^i_{2d}}{dt}&=k_1E^i_{1d}-k_2E^i_{2d}+u_dE^i_2-u_rE^i_{2d}\\
\frac{d I^i_d}{dt}&=k_2E^i_{2d}-qI^i_d-\frac{1}{D}I^i_d+u_dI^i-u_rI^i_{d}\\
\frac{d Q^i_d}{dt}&=qI^i_d-\frac{1}{D}Q^i_d+u_dQ^i-u_rQ^i_{d}\\
\frac{d R^i_d}{dt}&=\frac{1}{D}(I^i_d+Q^i_d)+u_dR^i-u_rR^i_{d}.
\end{split}
\end{align}

The equations for the physical distancing population mirrors (\ref{Model}) almost exactly. Firstly, we see that the signs of $u_d$ and $u_r$ are switched from \eqref{Model} because the physical distancing population exits \eqref{Model} and enters \eqref{SDModel}, and the returning to regular population exits \eqref{SDModel} and enters \eqref{Model}. Next, we see the modification that the susceptible population has a reduced rate of infection, as seen by the additional factor of $f^i(t)$ in the equations for $\frac{dS^i_d}{dt}$ and $\frac{d E^i_{1d}}{dt}$. We assume $f^i(t)$ to be a continuous, piecewise constant function with values in $[0,1]$, where $f^i(t)=1$ models zero reduction in contacts and $f^i(t)=0$ models $100\%$ reduction in contacts for physically distanced individuals. Moreover, the $f^i(t)$ remain constant during lockdown phases, with one week linear transitions between different phases. This results in,

\begin{align}
f^i(t)=\left\{\begin{array}{cc}
1, & t<t^F_{1}, \\
f^i_{2}+\frac{t^S_{2}-t}{t^S_{2}-t^F_{1}}\left(1-f^i_{2}\right), & t^F_{1} \leq t<t^S_{2}, \\
f^i_{2}, & t^S_{2} \leq t<t^F_{2}, \\
f^i_{3}+\frac{t^S_{3}-t}{t^S_{3}-t^F_{2}}\left(f^i_{2}-f^i_{3}\right), & t^F_{2} \leq t<t^S_{3}, \\
f^i_{3}, & t^S_{3} \leq t<t^F_{3}, \\
f^i_{4}+\frac{t^S_{4}-t}{t^S_{4}-t^F_{3}}\left(f^i_{3}-f^i_{4}\right), & t^F_{3} \leq t<t^S_{4}, \\
f^i_{4}, & t^S_{4} \leq t<t^F_{4}, \\
f^i_{5}+\frac{t^S_{5}-t}{t^S_{5}-t^F_{4}}\left(f^i_{4}-f^i_{5}\right), & t^F_{4} \leq t<t^S_{5}, \\
f^i_{5}, & t^S_{5} \leq t<t^F_{5}, \\
f^i_{6}+\frac{t^S_{6}-t}{t^S_{6}-t^F_{5}}\left(f^i_{5}-f^i_{6}\right), & t^F_{5} \leq t<t^S_{6}, \\
f^i_{6}, & t^S_{6} \leq t<t^F_{6}, \\
f^i_{7}+\frac{ t^S_{7}-t}{ t^S_{7}-t^F_{6}}\left(f^i_{6}-f^i_{7}\right), & t^F_{6} \leq t<t^S_{7}, \\
f^i_{7}, & t \geq t^S_{7},
\end{array}\right.\label{f}
\end{align}
where the dates for $t^S_j$ and $t^F_j$, for $j=1,\dots, 6$ are summarized in Table \ref{Phases}. 

These change points reflect the various phases of provincial lockdown within BC throughout 2020 as reported on \cite{BCCDC21a}. Specifically, Phase 1, Phase 2 and Phase 3A were between $t^F_1$ and $t^F_2$, $t^F_2$ and $t^F_3$, $t^F_3$ and $t^F_4$ respectively. According to our notation, Phase 3B would account of $t^F_4$ to $t^F_6$, and we included an additional change point for Thanksgiving at $t^F_5$. Finally, Phase 3C took effect after $t^F_6$.

\begin{table}[h!]
\centering
\begin{tabular}{|l|l|}
\hline
 Start Date  & End Date \\
\hline
\hline
  & $t^F_1=$ Mar. 14\\
\hline
 $t^S_2=$ Mar. 21 & $t^F_2=$ May 18\\
\hline
 $t^S_3=$ May 25 & $t^F_3=$ Jun. 23\\
\hline
 $t^S_4=$ Jun. 30 & $t^F_4=$ Sep. 12\\
\hline
 $t^S_5=$ Sep. 19 & $t^F_{5}=$ Oct. 12\\
\hline
 $t^S_{6}=$ Oct. 19 & $t^F_{6}=$ Nov. 7\\
\hline
$t^S_{7}=$ Nov. 14 & \\
\hline
\end{tabular}
\caption{The change points are used in the definition of $f^i(t)$. We note that all dates (except Thanksgiving) were obtained from \cite{BCCDC21a}.}\label{Phases}
\end{table}

As mentioned earlier in this section, the parameters $u_r$ and $u_d$ also determine the asymptotic proportion of the population following physical distancing protocols. This follows by defining $P^i(t)=S^i(t)+E^i_1(t)+E^i_2(t)+I^i(t)+R^i(t)+Q^i(t)$ and $P^i_d(t)=S^i_d(t)+E^i_{1d}(t)+E^i_{2d}(t)+I^i_{d}(t)+R^i_{d}(t)+Q^i_{d}(t)$, to represent the non-physically distanced population and the physically distanced population in the $i^{th}$ health region respectively. Using the systems (\ref{Model}) and (\ref{SDModel}) we obtain $\frac{d}{dt}P^i(t)+\frac{d}{dt}P^i_d(t)=0$, implying $P^i(t)+P^i_d(t)=N^i$, the total population of our modelled region. This is as expected, since we are not accounting for birth or death.  Taking $P^i(t)=N^i-P^i_d(t)$, we obtain the differential equation $\frac{d}{dt}P^i_d(t)=u_dN^i-(u_r+u_d)P^i(t)$, which implies that $P_d^{i}(t)=\frac{u_d}{u_r+u_d}N^i$ is a globally asymptotically stable fixed point of $\frac{d}{dt}P^i_d(t)$. Therefore, the system of differential equations in health region $i$, (\ref{Model}) and (\ref{SDModel}), rapidly converges to $\frac{u_d}{u_r+u_d}N^i$ individuals practising physical distancing. 

Throughout this paper, we will distinguish between the basic reproduction number in the absence of physical distancing, denoted by $R_{0b}$, and the regionalized basic reproduction number, denoted by $R^i_0$, which factors in physical distancing and quarantining. We obtain the equation for $R_{0b}$ by computing the basic reproduction number of a simplified version of (\ref{Model}), without physical distancing or quarantining. This can be done, for example, using the next-generation method of \cite{DHM90,VW02}, where $E^i_1, E^i_2$, and $I^i$ are taken as the disease states. This yields $R_{0b}=\beta (D+\frac{1}{k_2})$ which is common throughout the province, since $\beta, D$, and $k_2$ are not regionalized parameters. As shown in Appendix A of \cite{AndersonAE20}, the next-generation method can also be used to compute the regionalized basic reproduction number $R^i_0$, given by

\begin{align}
\begin{split}
R^i_0(t)=\beta \bigg( &\frac{e^4(1-e)(1-f^i(t))^2k_1k_2}{(e(\frac{1}{D}+q)+1)(ek_1+1)(ek_2+1)} +\frac{(ef^i(t)+1-e)^2}{\frac{1}{D}+q}\label{R0} \\
& + \frac{ek_1(ef^i(t)+1-e)^2}{k_2(ek_1+1)(ek_2+1)}+\frac{e(ef^i(t)+1-e)^2}{(ek_1+1)(ek_2+1)}\\ 
& + \frac{(ef^i(t)+1-e)^2}{k_2(ek_1+1)(ek_2+1)}+\frac{e^2k_1(e(f^i(t))^2+1-e)}{(ek_1+1)(ek_2+1)}\bigg),
\end{split}
\end{align}
where $e$ represents the asymptotic proportion of individuals practising physical distancing, given by $e=\frac{u_d}{u_r+u_d}$. Given that $R^i_0(t)$ depends on $f^i(t)$, it therefore varies in time as $f^i(t)$ transitions from $t^F_j$ to $t^S_{j+1}$ for $j=1,\dots 6$, and varies based on the different regions. 
\subsection{Statistical Modelling}\label{Stat Model}
With the epidemiological part of the model established, we now discuss the statistical aspects of our model. 

The ODEs of (\ref{Model}) and (\ref{SDModel}) are solved within Stan using a numerical ODE solver that includes arguments: $t$ (independent variable time), state (the ODE system at the time specified), $\theta$ (the ODE arguments that depend on parameters), $x_r$ (the ODE arguments that depend on data only) and $x_i$ (the integral data values used to evaluate the ODE system).

To relate the compartments of (\ref{Model}) and (\ref{SDModel}) to testing data, we follow the approach of \cite{AndersonAE20} and define the expected number of reported cases on day $r$ to be $\mu_r$, given by 
\begin{align} 
\mu^i_{r}=\psi^i(r) \int_{0}^{45} k_{2}\left[E^i_{2}(r-s)+E^i_{2 \mathrm{d}}(r-s)\right] w(s) \mathrm{d} s,
\label{eqn:mu_r}
\end{align}
where $\psi^i(r)$ is now the regional proportion of anticipated cases in the $i^{th}$ health region on day $r$ that have been tested and reported, and $w(s)$ is the density function of delay $s$, with a maximum delay of 45 days, with $w(\cdot)$ denoting the Weibull distribution. The histogram of times between symptoms onset and case reporting was fitted using a Weibull distribution, as in \cite{AndersonAE20} where the $99.99992\%$ quantile of 45 days is set to be the maximum delay. The resulting parameters for the Weibull distribution are $\lambda=9.85$ and $k=1.73$, which correspond to a mean of 8.8 days and SD 5.2 days.

Furthermore, given that testing protocol and availability evolved throughout the pandemic, we define $\psi^i(r)$ to be the piecewise function

\begin{align}
\psi^i(r)=\left\{\begin{array}{ccc}
\psi^i_1, &  &\text{March 1} \leq r \leq \text{March 15}, \\
\psi^i_2, &  &\text{March 16} \leq r \leq \text{April 8}, \\
\psi^i_3, &  &\text{April 9} \leq r \leq \text{April 20}, \\
\psi^i_4, &  &\text{April 21} \leq r \leq \text{December 31},
\end{array}\right.\label{psi}
\end{align}

We chose these change points for $\psi^i(r)$ because starting on March 16, 2020, the testing shifted focus to healthcare workers, long-term care residents, and community clusters not linked to travel. Then from April 9, 2020, expanded testing included residents in remote regions, people in homeless or unstable housing, first responders and returning travellers to Canada. Finally from April 21 and onwards, any individual with COVID-19 symptoms was eligible to get a test.

\begin{figure}[h]
\centering
\includegraphics[scale=0.8]{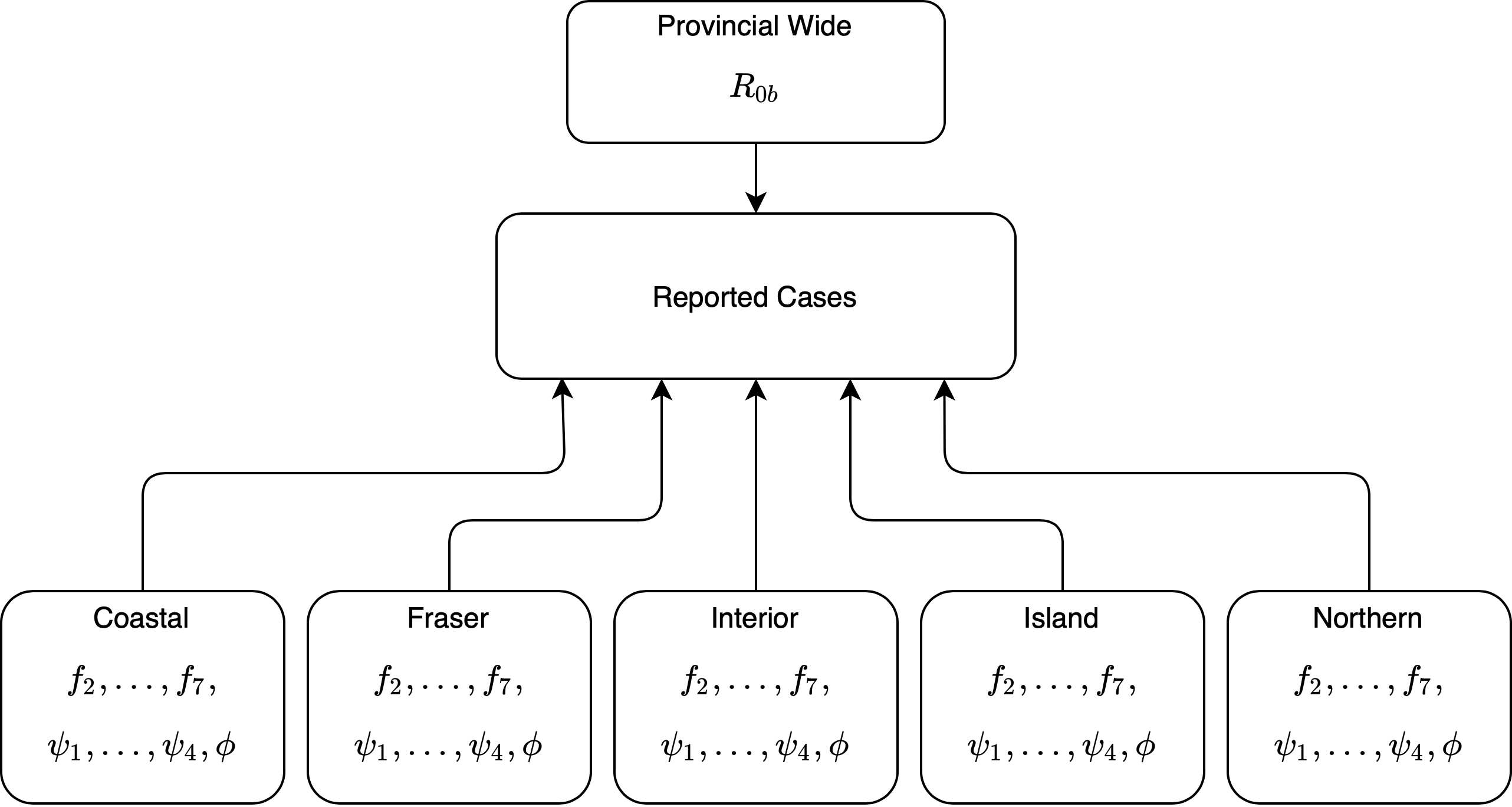}
\caption{This depicts the hierarchical structure of the model dependence on parameters. $R_{0b}$ is shared between all regions and $f_2,\dots,f_7$, $\psi_2,\dots,\psi_4$ and $\phi$ are individualized to each region.}
\label{fig:modelPic}
\end{figure}

As described earlier in Section \ref{Epi Model} and illustrated in Figure \ref{fig:modelPic}, we use a hierarchical model to have all five regions share a common $R_{0b}$, with the prior of Lognormal $(\log (2.6), 0.2)$. This prior reflects the range of reported values for the basic reproduction number of COVID-19 in early 2020 before large-scale interventions were implemented, as provided by \citet{midas}. 
This approach leads to the combining of regional and information from throughout BC to obtain smoothed estimates; that is the estimates are not purely driven by local dynamics.   The rate of spread within each region is then dependent on the regional specific parameters $f^i_2, \dots, f^i_7$.   We also expect $R_{0b}$ to be an important driver of the predictions made by the epidemiological model, 
with larger $R_{0b}$ resulting in an increased number of predicted cases. As discussed in Section \ref{Epi Model}, the parameters $D$ (the mean duration of the infectious period) and $k_2$ (rate of movement from $E^i_2$ to $I^i$) are fixed, thus ensuring that the dynamics of our model can clearly reflect a given value of $R_{0b}$, since $\beta=\frac{R_{0b}}{(D+1/k_2)}$.

For each region in the hierarchical model, we used a negative binomial observation model to link the expected case count in region $i$, $\mu^i_{r}$ from the structural model, and the observed case count data in region $i$, denoted by $C_{r}^i$:

\begin{equation}
\text{NB2}(C^i_r|\mu^i_r,\phi^i)={C^i_r+\phi^i-1 \choose C^i_r}\left(\frac{\mu^i_r}{\mu^i_r+\phi^i}\right)^{C^i_r}\left(\frac{\phi^i}{\mu^i_r+\phi^i}\right)^{\phi^i},
\label{eqn:nb}
\end{equation} 
where $\phi^i$ is the (inverse) dispersion parameter with prior $1/\phi^i \sim \chi^2_1$; this follows Stan's prior recommendations 
which guards against favouring a high over-dispersion \emph{a priori}. For more details, see \citet{stanpriors}. The negative binomial (with dispersion) is a commonly-used distribution to model random counts, when the variance is not necessarily a deterministic function of the mean. The separate dispersion parameters $\phi^i$ for each region allow for flexibility in modelling the negative binomial variance, which could differ between regions. Equation (\ref{eqn:nb}) makes the assumption that the observed cases in a region on a given day are conditionally independent of other days and regions, given its modelled mean and dispersion. Since $\mu^i_r$ is directly linked to the epidemiological model (see Equation \ref{eqn:mu_r}) and captures the trends in each region over time, we believe this conditional independence assumption is reasonable.  

Taking a Bayesian approach, the corresponding statistical model has parameters $R_{0 b}$, $f^i_{k}$, $\psi^i_j$ and $\phi^i$, for $k=2,\dots,7$ and $j=1, \dots, 4$, and the joint posterior distribution of the parameters can be written as

\begin{align}
\begin{split}
&\left[R_{0 b},\boldsymbol{f}_2,\dots, \boldsymbol{f}_7, \boldsymbol{\psi}_{1},\dots, \boldsymbol{\psi}_{4}, \boldsymbol{\phi} \mid \boldsymbol{C}\right] \propto\\
& \hspace{4mm} \left[R_{0b}\right] \prod_{i=1}^5  \left\{ \left[\boldsymbol{C}^i \mid R_{0 b},f^i_2,\dots, f^i_7, \psi^i_{1},\dots, \psi^i_{4}, \phi^i\right]\left[f^i_{2}\right]\dots\left[f^i_{7}\right]\left[\psi^i_{1}\right]\dots\left[\psi^i_{4}\right][\phi^i] \right\},
\end{split}\label{eqn:post}
\end{align}
where $\left[\boldsymbol{C}^i \mid R_{0 b},f^i_2,\dots, f^i_7, \psi^i_{1},\dots, \psi^i_{4}, \phi^i\right]$ represents the negative binomial data likelihood for region $i$, the prior distributions are shown by $\left[R_{0b}\right], \left[f^i_{2}\right], \dots ,\left[f^i_{7}\right], \left[\psi^i_{1}\right], \dots, \left[\psi^i_{4}\right]$ and $[\phi^i]$, and $\boldsymbol{C}^i$ is the vector of reported cases for each region. The prior distributions of $\left[R_{0b}\right], \left[f^i_{2}\right], \dots ,\left[f^i_{7}\right]$, and $\left[\psi^i_{1}\right], \dots, \left[\psi^i_{4}\right]$ are summarized in Table \ref{Priors}.
The priors chosen for $f^i_{2}\dots f^i_{7}$ reflect the gradual loosening of restriction between March 21, 2020 and November 6, 2020, and tightening of restrictions starting from November 7, 2020. Equation (\ref{eqn:post}) further indicates that we assume the priors are independent. Thus, for example, we expect \emph{a priori} that $f_3^i$ is likely to be larger than $f_2^i$ due to gradual loosening of restrictions and so Table \ref{Priors} places their prior modes at 0.5 vs.~0.4; however, this choice of prior does not force $f_3^i$ to be larger than $f_2^i$ in the posterior distribution due to their independence. For $\psi_1, \ldots, \psi_4$, the prior modes reflect our knowledge of increasing test availability, but these quantities cannot be directly measured; the antibody study of \cite{antibody} provides a rough estimate of antibody seroprevalence due to infection and it is reasonable to set the prior mode for $\psi_4$ (which covers the majority of the study period duration) at 0.4 with a wide SD of 0.2, given the high degree of uncertainty in the Statistics Canada estimate.\footnote{A 95\% CI of $(0.5, 2.9)$ is reported as the percentage of BC population having antibody seroprevalence due to COVID-19 infection, based on survey data collected from November 2020 to April 2021. This may be compared to the 52,817 total cases observed in BC through December 2020, representing 1.04\% of the population.} 
In general, the priors we have chosen encode our knowledge of policy and testing protocols, while being flexible enough (via prior standard deviations of 0.25 for $f^i_{2}\dots f^i_{7}$ and 0.2 for $\psi^i_{1},\dots, \psi^i_{4}$) to allow the data to inform the final parameter estimates. This point is validated in the simulation study that follows in Section 2.4.

\begin{table}[h!]
\centering
\begin{tabular}{|c|c|c|c|}
\hline
 Parameter & Prior Distribution &  Mode & Standard Deviation\\
\hline
\hline
 $R_{0b}$ &  Lognormal $(\log (2.6), 0.2)$ & 2.6 & 0.2871\\
\hline
 $f^i_2$ & $\operatorname{Beta}(1.393,1.590)$ & 0.4 & 0.25\\
 $f^i_3$ & $\operatorname{Beta}(1.500,1.500)$ & 0.5 & 0.25\\
 $f^i_4$ & $ \operatorname{Beta}(1.590,1.393)$ & 0.6 & 0.25\\
 $f^i_5$ & $\operatorname{Beta}(1.655,1.281)$ & 0.7 & 0.25\\
 $f^i_6$ & $ \operatorname{Beta}(1.694,1.174)$ & 0.8 & 0.25\\
 $f^i_7$ & $\operatorname{Beta}(1.590,1.393)$ & 0.6 & 0.25\\
\hline
$\psi^i_1$ & $\operatorname{Beta}(1.217, 2.951)$ & 0.1 & 0.2\\
$\psi^i_2$ & $\operatorname{Beta}(1.509, 3.036)$ & 0.2 & 0.2\\
$\psi^i_3$ & $\operatorname{Beta}(1.870, 3.030)$ & 0.3 & 0.2\\
$\psi^i_4$ & $\operatorname{Beta}(2.263, 2.894)$ & 0.4 & 0.2\\
\hline
\end{tabular}
\caption{\label{Priors} The prior distributions for the parameters in the statistical model.}
\end{table}

\subsection{Provincial and Regional Model Inputs}
\label{sec:modelInput}
In this section, we summarize the rest of the model inputs from known sources.

Table \ref{table:pop} shows the populations of each region reported in \cite{InteriorHealth20}. This was used for initialization and comparison with a provincial-wide model. Moreover, we have used the value of $N=$5,100,000 in our provincial-wide model for the population of BC\footnote{Although the sum of the regional population is 5050481, we have used the same provincial-wide population as \cite{AndersonAE20} for consistency.}.

\begin{table}[h!]
\centering
\begin{tabular}{|c|c|}
\hline
Region & Population\\
\hline
\hline
Coastal & 1,225,195\\
\hline
Fraser & 1,889,225\\
\hline
Interior & 795,116\\
\hline
Island & 843,375\\
\hline
Northern & 297,570\\
\hline
\end{tabular}
\caption{Populations of each regional Health Authority in BC.}
\label{table:pop}
\end{table}

For the provincial-wide model, we treat the entire province as one region and thus have a single set of differential equations \eqref{Model} and \eqref{SDModel} to solve. We take the same initial conditions as those given in \cite{AndersonAE20}. Specifically, the model is initialized on February 1 with 8 individuals shared between  compartments $E_1, E_2, I$ and $E_{1d}, E_{2d}$ and $I_d$, where the 8 cases reflects an assumed 10-30\% reporting rate in the early days of the pandemic (see Table 2 of \cite{AndersonAE20} for the specific values). Using these initial conditions, equations (\ref{Model}) and (\ref{SDModel}) are solved throughout February to the obtain compartment values needed for the observation model beginning on March 1. The purpose of including the provincial-wide model is to subsequently compare results with those of our regional model. To do so, we impose the same model structure on the provincial-wide model, with  the same change points for $f_2, \dots, f_7$ and $\psi_1,\dots, \psi_4$ and the same priors on all model parameters. Then to fairly compare the results with those from the regional model, we scale our provincial-wide results using the regional population ratios ($\frac{N^i}{N}$), i.e., Coastal by $0.24$, Fraser by $0.37$, Interior by $0.16$, Island by $0.17$ and Northern by $0.06$.

For the regional model, we have equations (\ref{Model}) and (\ref{SDModel}) to solve for each region given the set of parameters $f^i_1,\dots, f^i_7$ and $R_{0b}$, or equivalently $\beta$. Likewise, appropriate initial conditions for the differential equations are needed, namely initial values for all model compartments in all five regions. To obtain these, we similarly scale the provincial-wide compartmental values on February 1 according to their relative population. For example, in the $i$-th region we set $S^i(\text{February 1})=\frac{N^i}{N}S(\text{February 1})$, where $S$ denotes total number of susceptible individuals provincial-wide and $N$ is the total population of BC.

The data used in this research are publicly available from the British Columbia Centre for Disease Control website of \cite{BCCDC21a}. At the time that this project commenced, the data for all health authorities were downloaded in a single CSV file and was manually organized by region. During the time frame under investigation in this work, BC experienced two waves of COVID-19 cases. The first occurred from March to early April of 2020, where the seven-day moving average of cases (red line) reached approximately 50 cases per day. From the beginning of April onward, case numbers were very low and remained that way until they started climbing at the beginning of July 2020. The second wave surge commenced at the end of October 2020. The seven-day moving average of cases skyrocketed from 150 cases per day to just under 800 cases per day by the beginning of December 2020 before tapering off and plateauing at approximately 500 cases per day for the remainder of 2020. 

\subsection{Simulation Study}\label{sec:simstudy}

In this section, we discuss details on validating our modelling approach using simulated data.

We conducted a simulation study to assess the estimation of model parameters in a controlled setting. Starting with known values of all the parameters, we use the model to simulate observations (case counts). The model is then fitted to the simulated data, and we examine whether the original parameter values can be recovered.  We follow the model setup as laid out in the previous sections, including the priors (Table \ref{Priors}), change-points for $f^i(t)$ (Table \ref{Phases} and transitions in Equation \ref{f}) and  $\psi^i(r)$ (Equation \ref{psi}), and initialization of the compartments (Section \ref{sec:modelInput}).

The values set for the hierarchical $R_{0 b}$ and regional $\left\{ f^i_2,\dots, f^i_7, \psi^i_{1},\dots, \psi^i_{4}, \phi^i \right\}_{i=1}^5$ are displayed in Table \ref{tab:simstudy}; these values were chosen such that case counts simulated from these parameters can mimic the real data observed for the five regions. In particular, there is substantial heterogeneity in several of the regional parameters (e.g., $f_3, \psi_1, \psi_2, \phi$), and the modes of the priors are not necessarily close to the actual parameter values (e.g., hierarchical $R_{0b}$, $f_3$ and $f_5$ for most regions, $\psi(r)$ for Coastal region). Thus, this is also a realistic test for whether the guidance provided by the priors leave sufficient flexibility for the data to inform the parameter estimates.

\begin{table}[ht]
\centering
\begin{tabular}{|c|c|c|c|c|c|}
\hline
  \multirow{2}{*}{Parameter} & \multicolumn{5}{c|}{Region} \\
 \cline{2-6}  
 & Coastal & Fraser & Interior & Island & Northern \\ 
  \hline
  \hline
$R_{0b}$& \multicolumn{5}{c|}{3.00 (common to all regions) } \\ 
  \hline
  $f_2$ & 0.33 & 0.42 & 0.20 & 0.16 & 0.32 \\ 
  $f_3$ & 0.72 & 0.59 & 0.95 & 0.79 & 0.66 \\ 
  $f_4$ & 0.66 & 0.63 & 0.52 & 0.62 & 0.67 \\ 
  $f_5$ & 0.46 & 0.63 & 0.68 & 0.45 & 0.39 \\ 
  $f_6$ & 0.79 & 0.75 & 0.79 & 0.99 & 0.87 \\ 
  $f_7$ & 0.49 & 0.52 & 0.62 & 0.49 & 0.64 \\ 
    \hline
  $\psi_1$ & 0.39 & 0.11 & 0.02 & 0.06 & 0.07 \\ 
  $\psi_2$ & 0.42 & 0.22 & 0.15 & 0.10 & 0.07 \\ 
  $\psi_3$ & 0.38 & 0.26 & 0.24 & 0.27 & 0.22 \\ 
  $\psi_4$ & 0.66 & 0.60 & 0.70 & 0.54 & 0.44 \\ 
  \hline
  $\phi$ & 8.00 & 11.00 & 3.00 & 8.00 & 5.00 \\ 
   \hline
\end{tabular}
\caption{Parameter values for simulation study.} \label{tab:simstudy}
\end{table}

To simulate observations from the model, we use the initial conditions (Section \ref{sec:modelInput}) to solve the ODE system  (Equations \ref{Model} and \ref{SDModel}) and obtain compartment values for each region from February 1 to December 31 of 2020. Then we use Equation (\ref{eqn:mu_r}) to calculate the expected number of reported cases for each region on each day from March 1 to December 31 of 2020, and finally generate our simulated daily case counts from the negative binomial observation model (Equation \ref{eqn:nb}).

We then fit the model to the simulated data, and the histograms of the MCMC samples from Stan for each parameter are shown in Figures \ref{fig:simhist-regional-a}--\ref{fig:simhist-regional-c} in Appendix \ref{sec:simstudyplot}, with the blue vertical lines indicating the true value of the parameter (as given in Table \ref{tab:simstudy}) and the red lines indicating the 0.05 and 0.95 quantiles (i.e., the bounds of the 90\% central credible interval) of the MCMC samples for that parameter. It can be seen that all of the true parameter values lie within their 90\% credible intervals, which suggests that the Bayesian estimation procedure can reasonably infer the model parameters given the choice of priors. For example, the prior mode for the hierarchical $R_{0b}$ is at 2.6, while its posterior distribution (last panel of Figure \ref{fig:simhist-regional-c}) has shifted to a mode just below the true value of 3.0 together with a fairly precise 90\% credible interval (2.90, 3.08). The posteriors of $f_i$ likewise have fairly tight credible intervals, showing that the estimates are not unduly influenced by the priors (e.g., see the $f_3$ panels in the second column of Figure \ref{fig:simhist-regional-b}, which clearly reflect the differences in the true regional $f_3$ values in Table \ref{tab:simstudy}). The most uncertain parameter is $\psi_4$, as seen in the relatively wide credible intervals (Figure \ref{fig:simhist-regional-a}, fourth column); nonetheless, a shift away from the prior mode (0.4) towards higher values is noticeable. Overall, these results provide confidence in proceeding to apply our hierarchical model to the real data analysis.

\subsection{Justification of the Hierarchical Regional Model}\label{sec:Just}
In this section, we further justify our choice of a hierarchical regional model with $\beta$ being a provincial-wide estimated parameter and $f^i(t)$ being regional estimated parameters. As discussed at the beginning of Section \ref{Epi Model}, there is dependence between the rate of infection parameter $\beta$ and the relative reduction in physical contacts $f^i(t)$. In particular, we stated earlier in Section \ref{Epi Model} that the same data can be fitted by having decreases in $\beta$ compensated by increases in $f^i(t)$, or vice versa. This inverse relationship can be seen analytically using equation \eqref{R0} for the basic reproduction number $R_0^i$ for region $i$. Using the known parameters defined in Table \ref{table:constants}, equation \eqref{R0} simplifies to 
\begin{equation}
R_0^i(t)=\beta\left(0.151932+1.3628f^i(t)+3.48527({f^{i}}(t))^2\right),\label{Just:R0}
\end{equation}
with $R_{0b}=\beta(D+\frac{1}{k_2})=6\beta$. 

Thus, we see that indeed increases or reductions in $\beta$, or equivalently in $R_{0b}$, can be compensated for by reductions or increases respectively in $f^i(t)$ to achieve the same $R_0^i$. 
This inverse relationship is also confirmed in Figures \ref{fig:R0bCompIndiv} and \ref{fig:fCompIndiv} of Appendix \ref{sec:NHregional}. In these plots, we are comparing a non-hierarchical regional model, meaning both $\beta$ and the $f^i$s are estimated using only regional data, versus our hierarchical regional model with an estimated $\beta$ taken to be the common throughout the province. Specifically in Figure \ref{fig:R0bCompIndiv}, we see that the non-hierarchical regional model estimates $R_{0b}$, shown in blue, to be consistently lower than the provincial estimates, with more pronounced variation in the less populated regions of Island and Northern. Looking at Figure \ref{fig:fCompIndiv}, we also see that the non-hierarchical regional model, shown in blue, estimates consistently larger $f^i(t)$ than the hierarchical regional model. Therefore, although the non-hierarchical regional model may be able to obtain an adequate fit of the data, the increased variability in $R_{0b}$ throughout the regions inhibits our ability to interpret the estimated $f^i(t)$ values. Moreover, the stability of the parameter estimates is an important feature of the hierarchical regional model, as validated in the simulation study (Section \ref{sec:simstudy}) in which the original parameter values could be largely recovered from sample data. Furthermore, we again emphasize that the priors chosen for the hierarchical regional model provide adequate constraints on the parameters, especially $R_{0b}$ and $f^i(t)$, while allowing the data to inform the final parameter estimates.

\section{Results and Discussion} \label{results}

In this section, we present numerical results which highlight the differences between provincial and regional modelling. We focus on two comparisons, the first being a full analysis using reported cases from March 1 to December 31 of 2020, comparing the total number of active cases or prevalence inferred from the fitted provincial-wide and regional models. We emphasize that the prevalence is different from the total number of active cases which have been reported.  Prevalence takes into account all active cases within a specific region, not just those who have tested positive. The second comparison is to illustrate the predictive ability of the fitted provincial-wide and regional models, where we use the data from March 1 to November 7 of 2020 and let both models predict the future prevalence. Overall both comparisons show significant differences between the provincial-wide and regional models. Below, we present a detailed discussion of these results and provide some region-specific recommendations based on model predictions.  To obtain the samples from the posterior distribution of the parameters, we used Stan 2.21.0 and R 4.0.2, running 2000 HMC iterations and 4 chains. 
All 4 chains were observed to have converged, as shown in the trace plots of Figures \ref{fig:traceplots-regional-a}-\ref{fig:traceplots-BCwide} from Appendix \ref{sec:traceplot}.

\subsection{Comparison between Provincial-wide and Hierarchical Regional Models}
\label{sec:prevalence}

\begin{figure}[h!]
\centering
\includegraphics[scale=0.9]{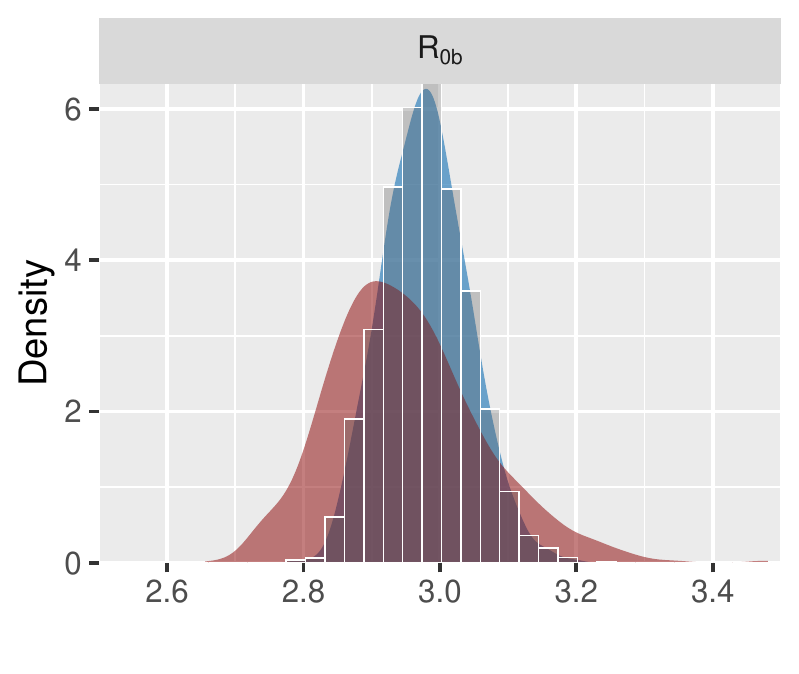}
\vspace{-6mm}
\caption{$R_{0b}$ comparison of hierarchical regional model (blue) vs provincial-wide model (red).}
\label{fig:R0b}
\end{figure}

\begin{figure}[h!]
 \vspace{-0.4in}
    \centering
    \begin{subfigure}[b]{\textwidth}
        \centering
        \includegraphics[scale=0.7]{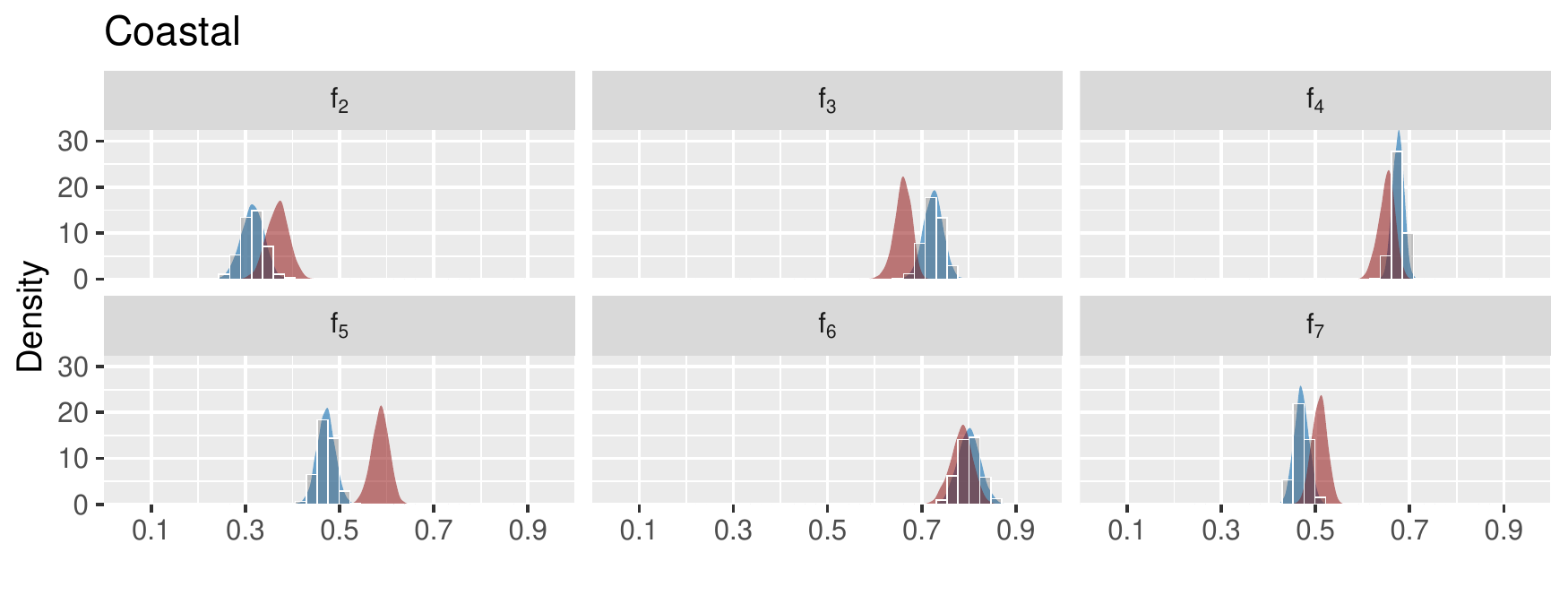}
    \end{subfigure}\\
    \begin{subfigure}[b]{\textwidth}
        \centering
        \includegraphics[scale=0.7]{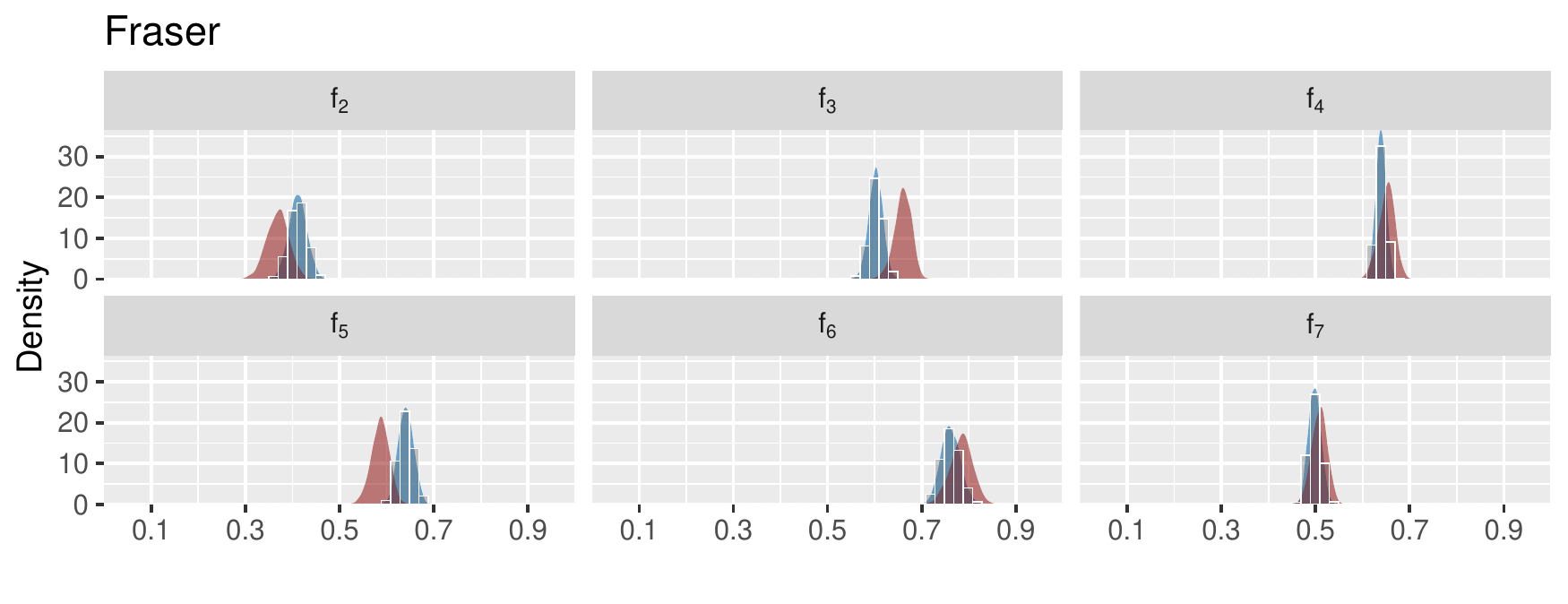}
    \end{subfigure}\\
    \begin{subfigure}[b]{\textwidth}
        \centering
        \includegraphics[scale=0.7]{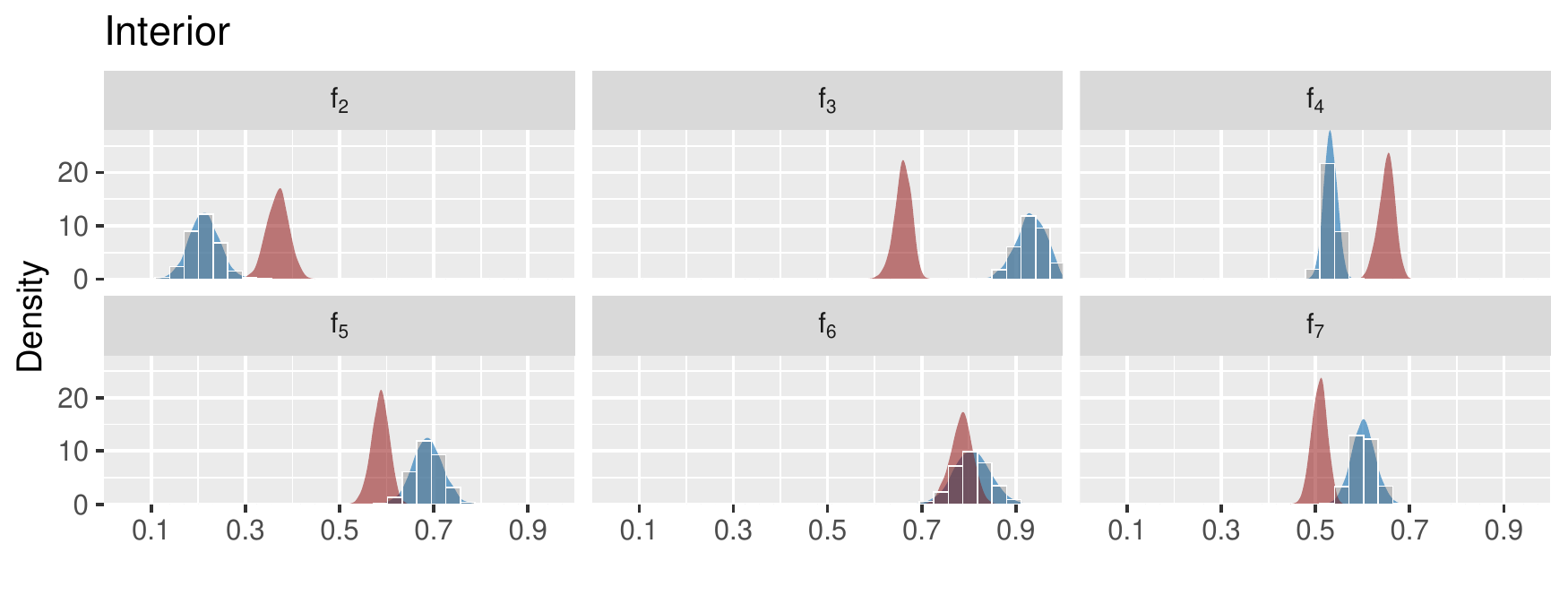}
    \end{subfigure}\\
    \begin{subfigure}[b]{\textwidth}
        \centering
        \includegraphics[scale=0.7]{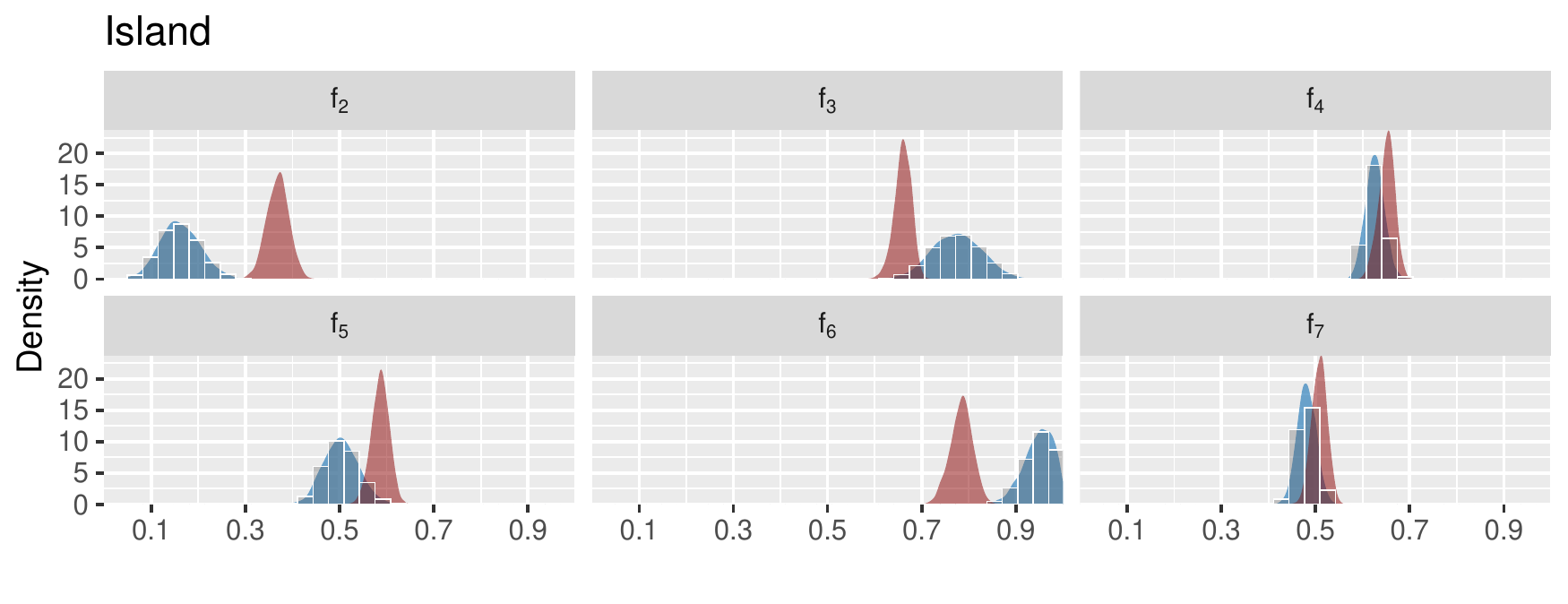}
    \end{subfigure}\\
    \begin{subfigure}[b]{\textwidth}
        \centering
        \includegraphics[scale=0.7]{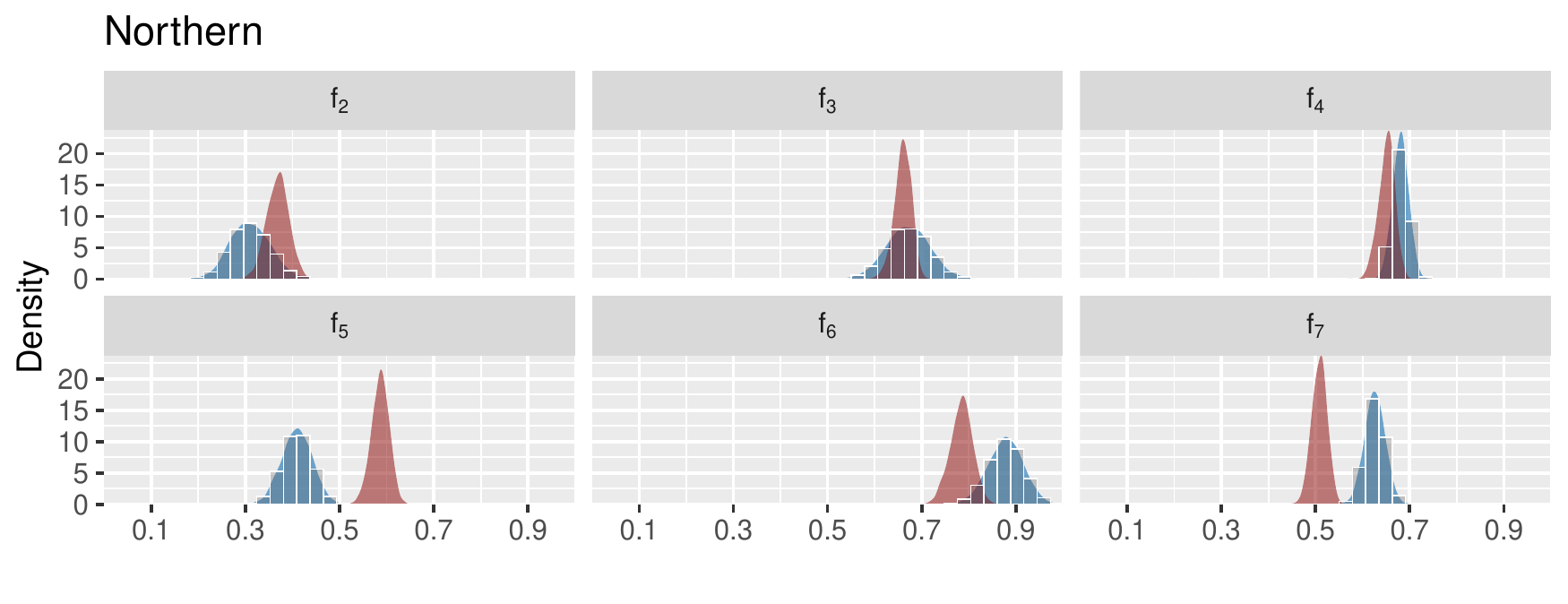}
    \end{subfigure}
    \vspace{-10mm}
    \caption{Comparison of estimated posterior probability densities for the relative reduction in contact due to physical distancing, $f_2,\dots,f_7$ by regions (blue) and provincial-wide (red).  Red and blue densities with little overlap indicate that the provincial-wide estimates are substantively different from those of the regional model.}
    \label{fig:fComp}
\end{figure}

\begin{figure}[h!]
    \centering
    \begin{subfigure}[b]{\textwidth}
        \centering
        \includegraphics[scale=0.6]{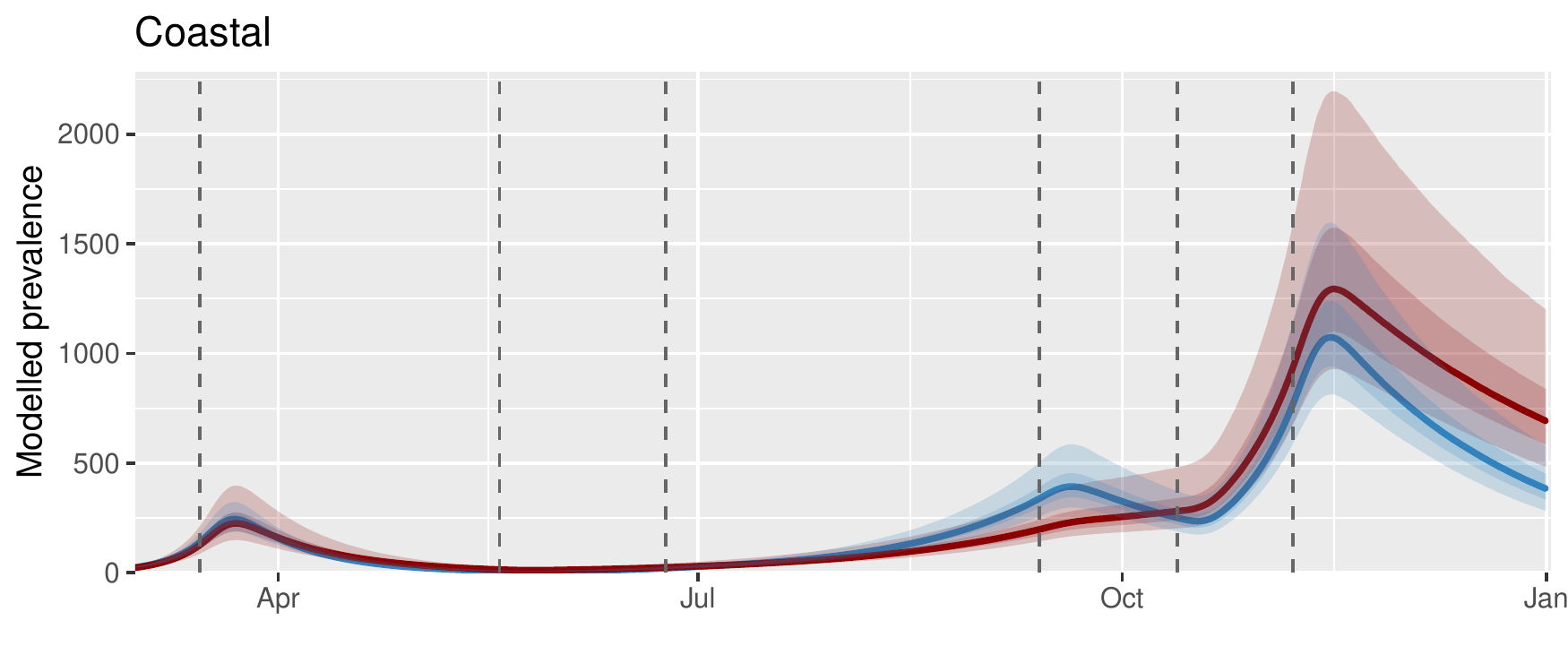}
    \end{subfigure}\\
    \begin{subfigure}[b]{\textwidth}
        \centering
        \includegraphics[scale=0.6]{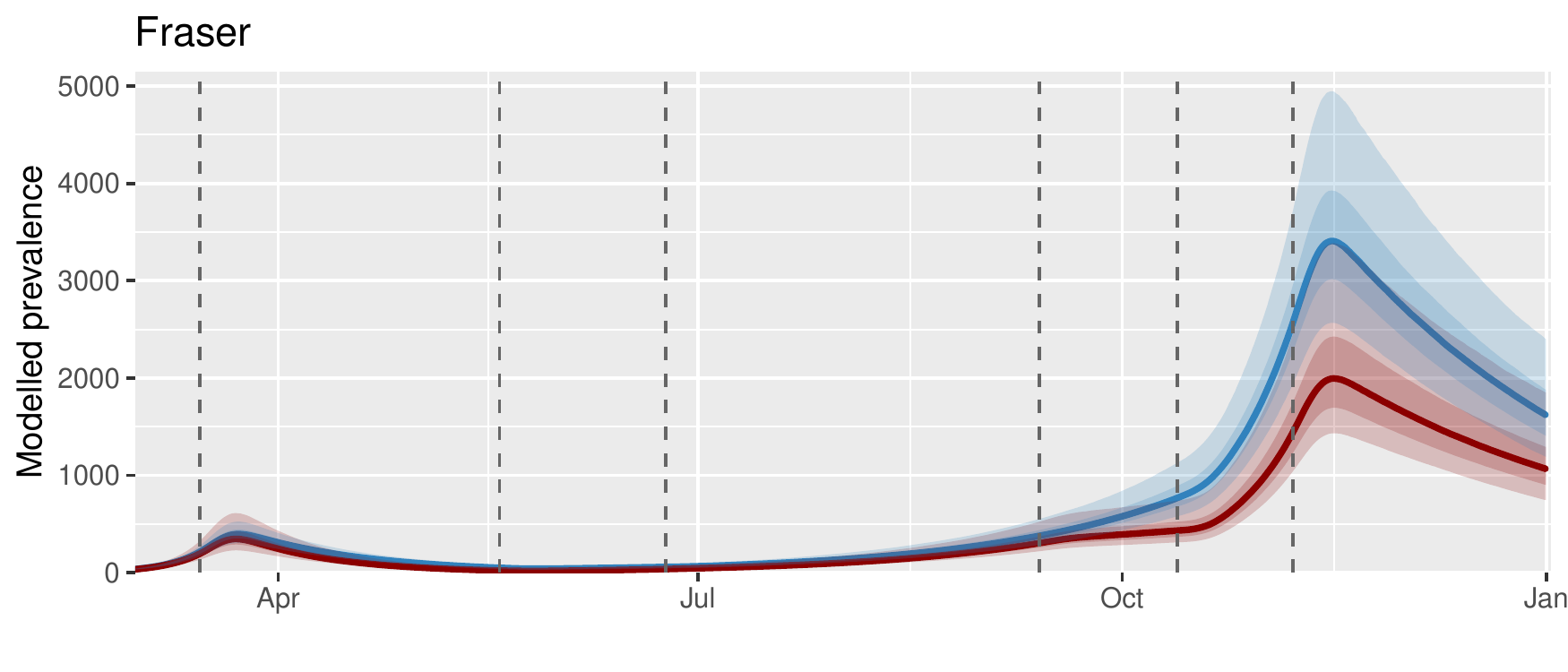}
    \end{subfigure}\\
    \begin{subfigure}[b]{\textwidth}
        \centering
        \includegraphics[scale=0.6]{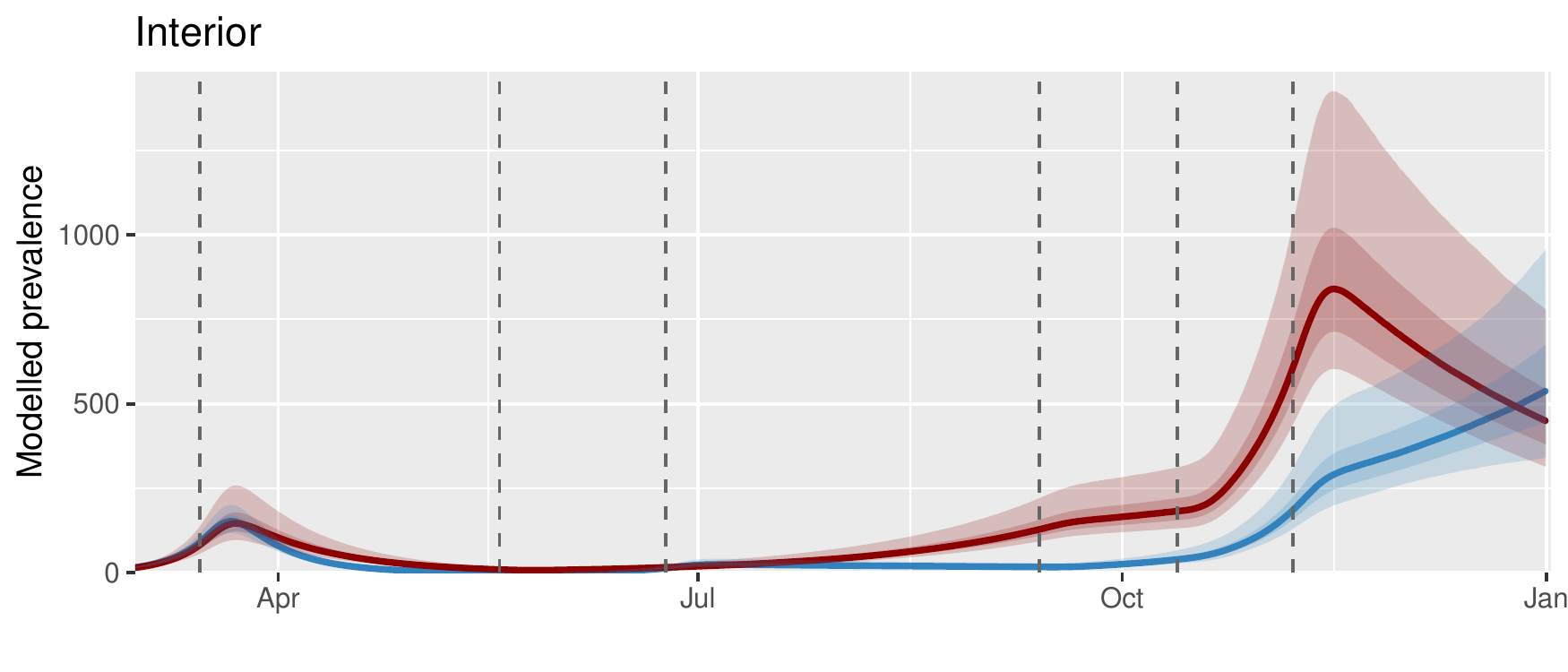}
    \end{subfigure}\\
    \begin{subfigure}[b]{\textwidth}
        \centering
        \includegraphics[scale=0.6]{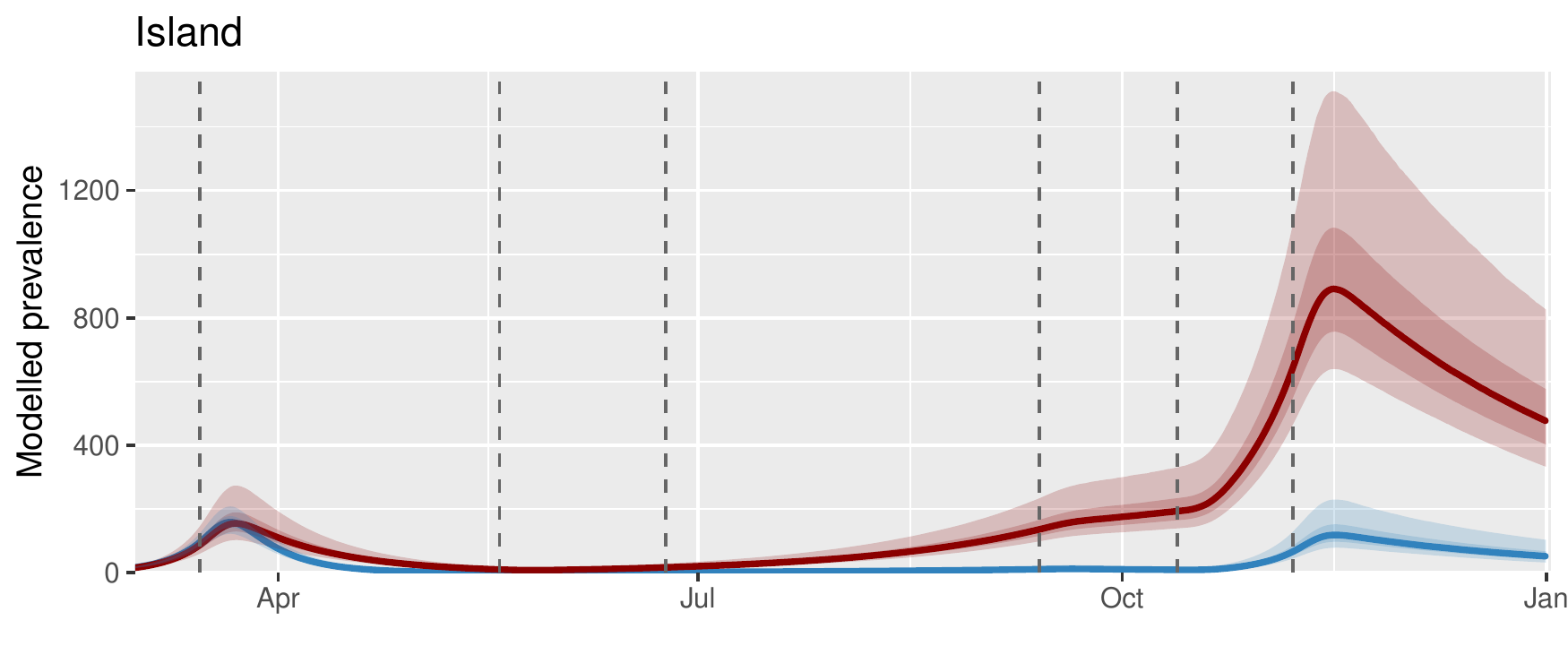}
    \end{subfigure}\\
    \begin{subfigure}[b]{\textwidth}
        \centering
        \includegraphics[scale=0.6]{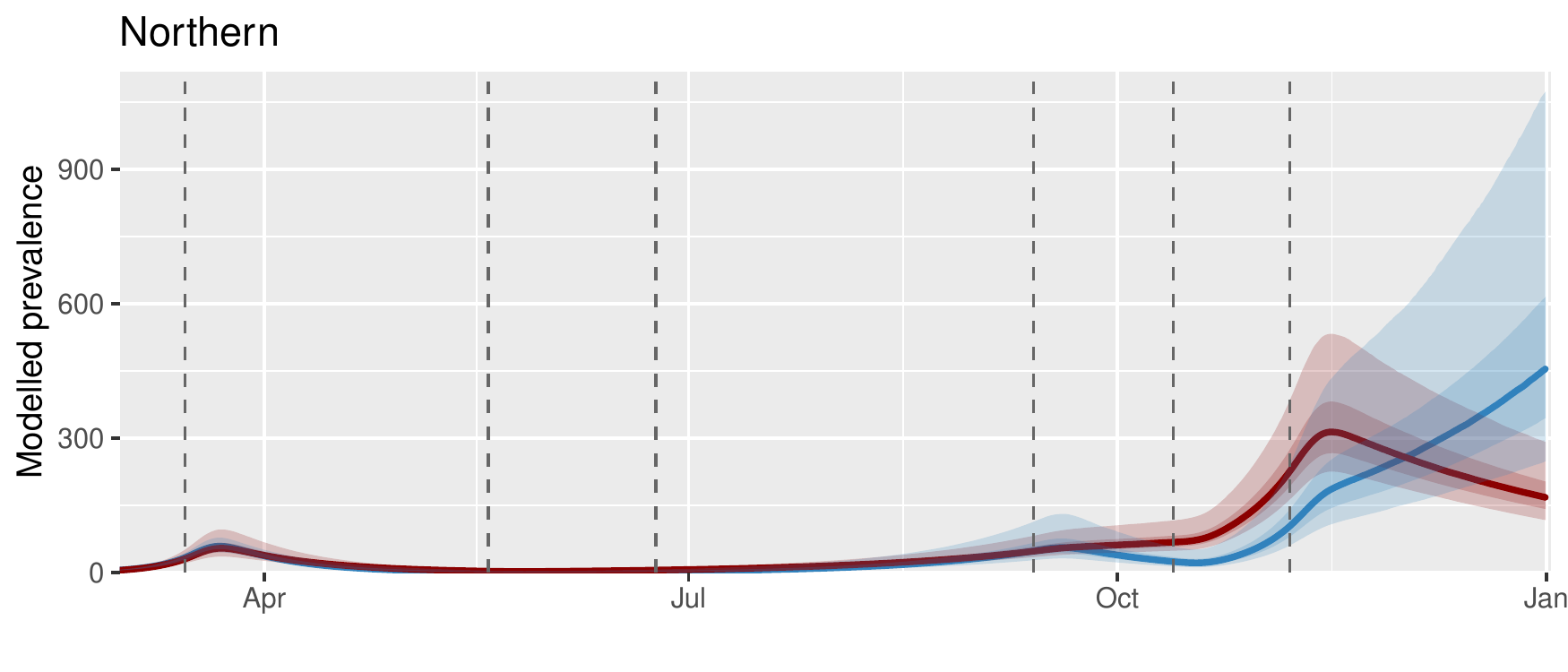}
    \end{subfigure}
    \vspace{-10mm}
    \caption{Comparison of modelled prevalence by regions (blue) and provincial-wide (red) scaled by respective regional population ratios. The dotted lines indicate start of reopening phases of $f_2^i,\dots,f_7^i$, as detailed in Table \ref{Phases}. Here, the solid lines indicate the respective mean and shaded regions indicate respective ranges with 50\% and 90\% posterior probability.}
    \label{fig:prev}
\end{figure}

\begin{figure}[h!]
    \centering
    \begin{subfigure}[b]{\textwidth}
        \centering
        \includegraphics[scale=0.6]{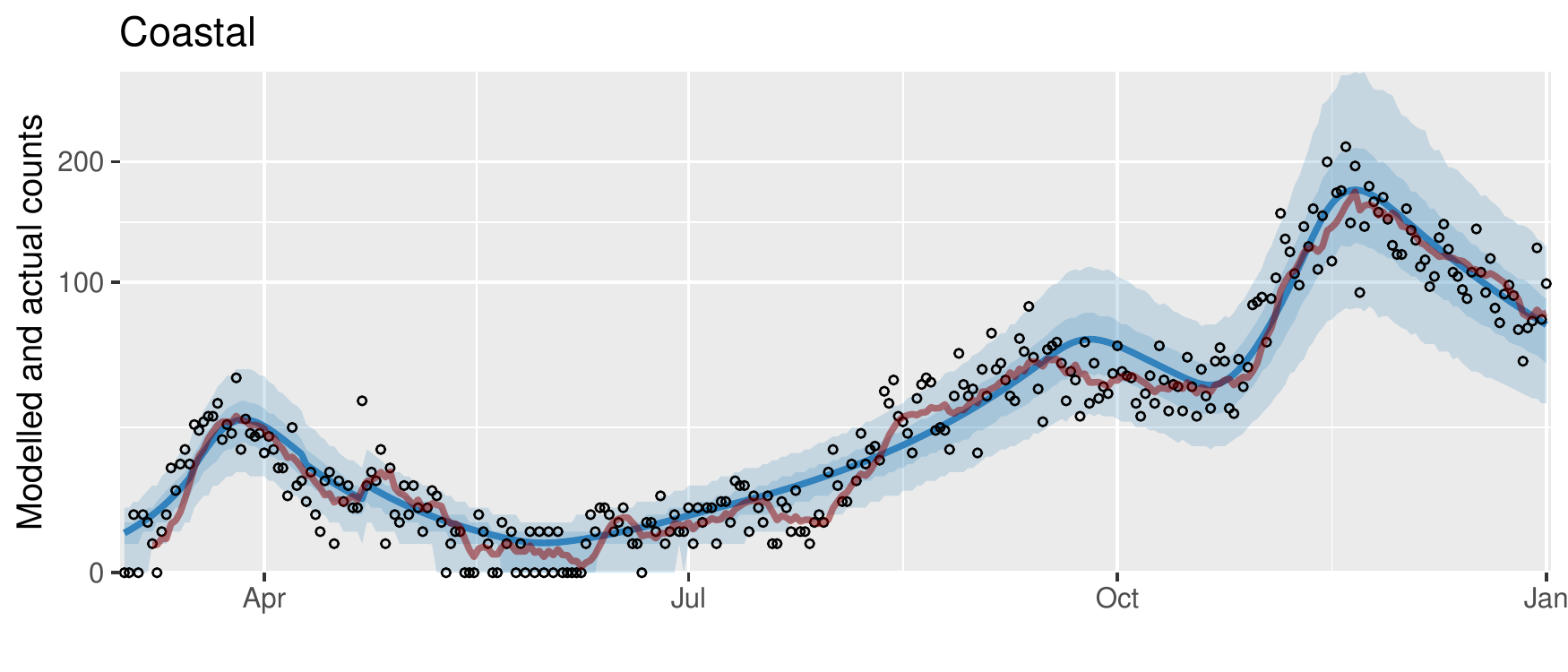}
    \end{subfigure}\\
    \begin{subfigure}[b]{\textwidth}
        \centering
        \includegraphics[scale=0.6]{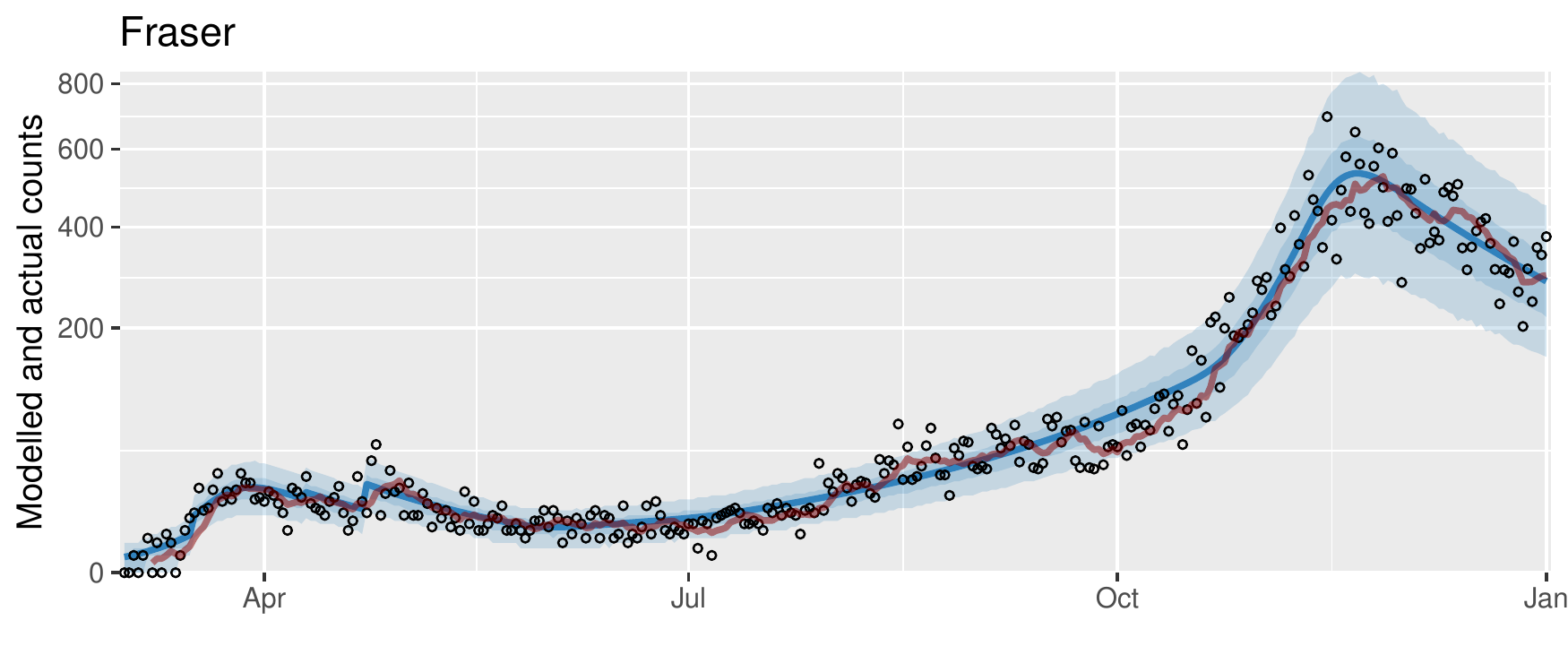}
    \end{subfigure}\\
    \begin{subfigure}[b]{\textwidth}
        \centering
        \includegraphics[scale=0.6]{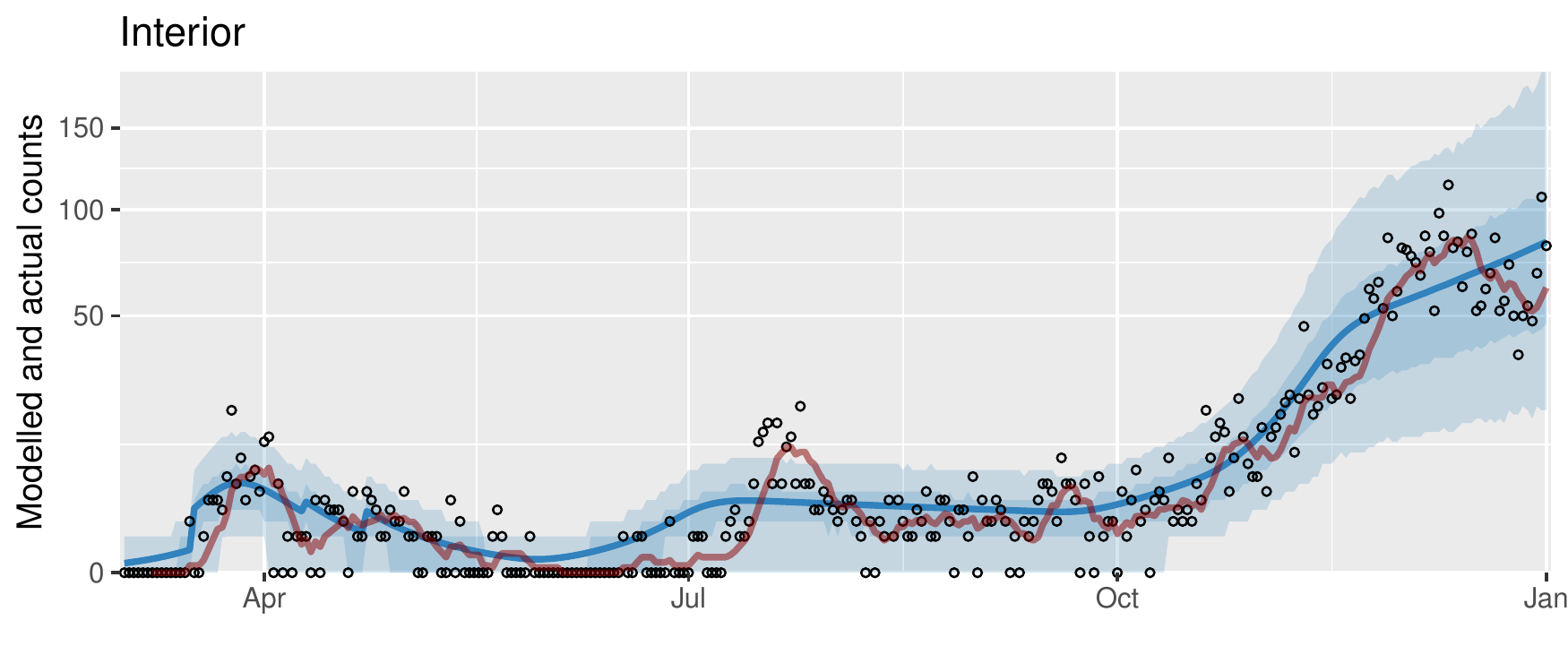}
    \end{subfigure}\\
    \begin{subfigure}[b]{\textwidth}
        \centering
        \includegraphics[scale=0.6]{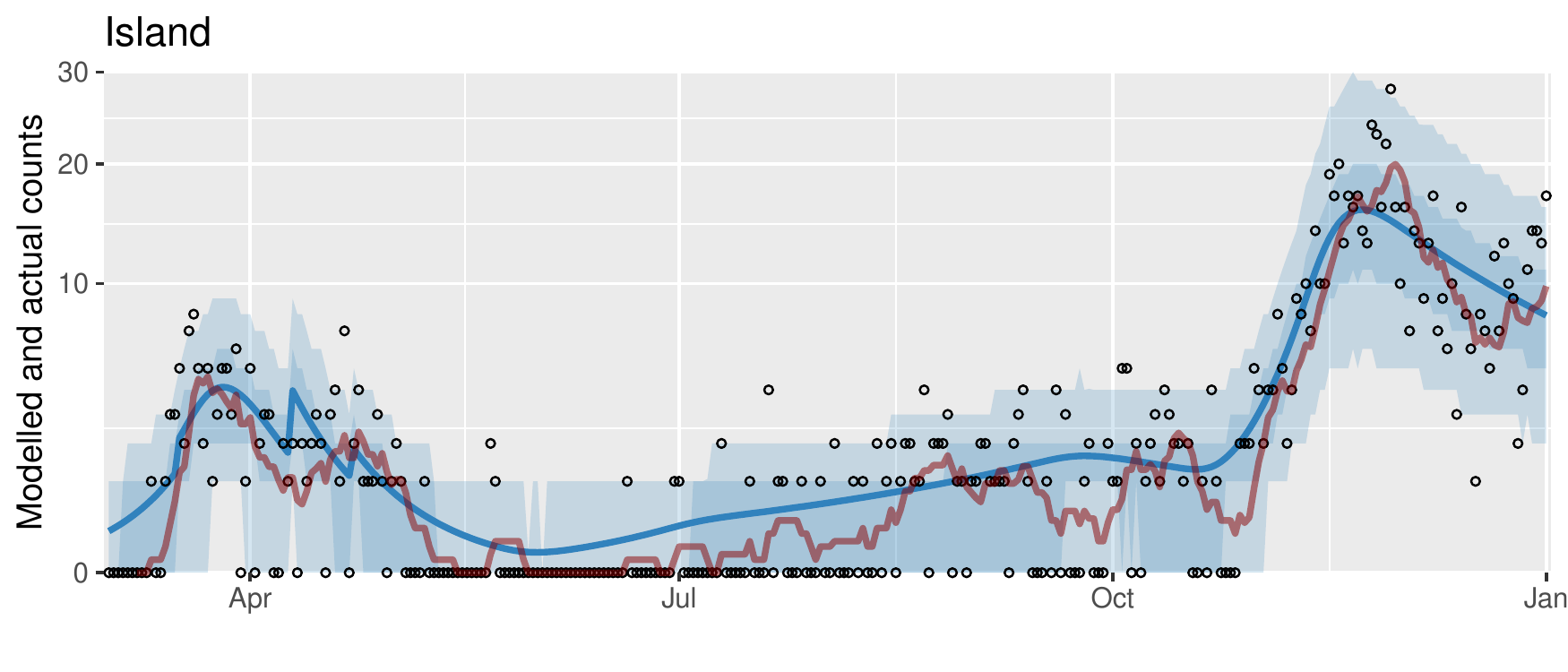}
    \end{subfigure}\\
    \begin{subfigure}[b]{\textwidth}
        \centering
        \includegraphics[scale=0.6]{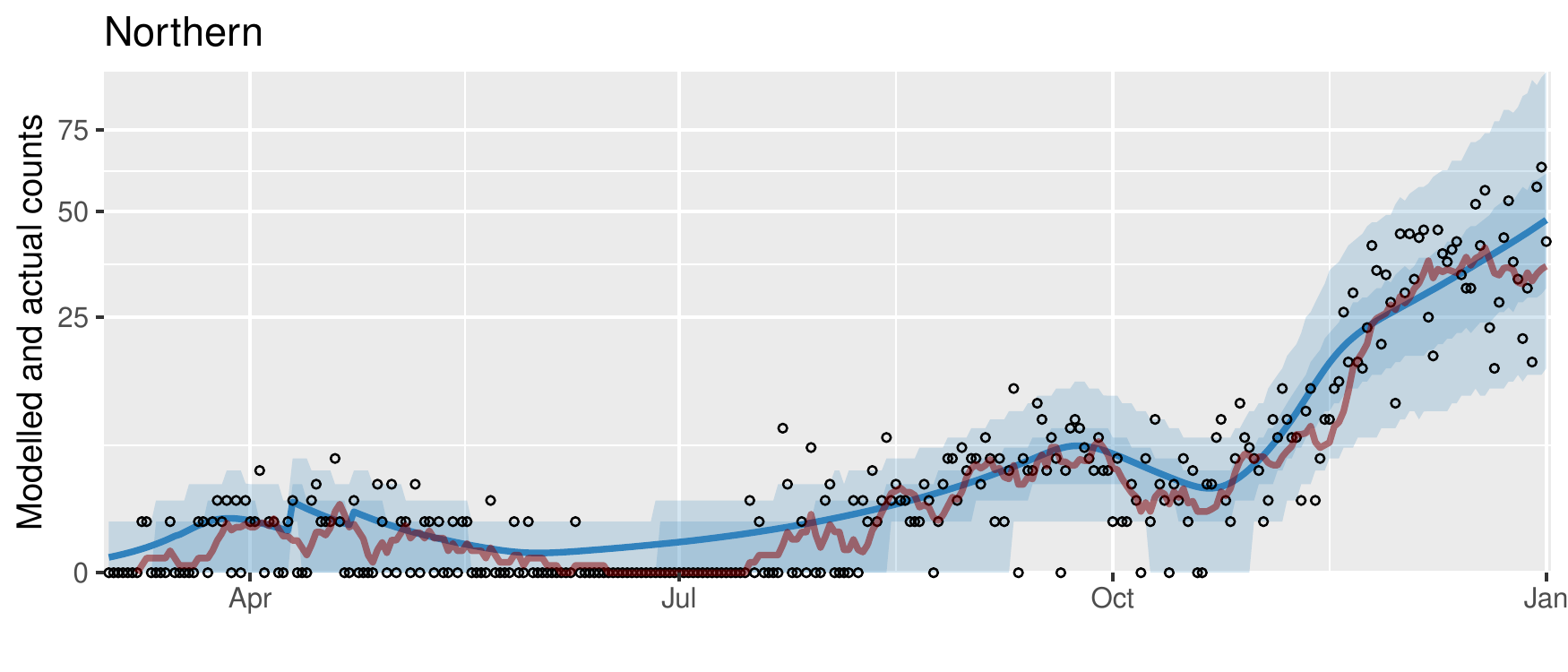}
    \end{subfigure}
    \vspace{-10mm}
    \caption{Comparison on reported (dots) and modelled  cases (blue) of each region. Solid blue line indicate the mean and blue shaded regions indicate the ranges of modelled cases with 50\% and 90\% posterior probability. The 7-day moving average of case counts is shown in red.}
    \label{fig:count}
\end{figure}

\begin{figure}[h!]
    \centering
    \begin{subfigure}[b]{.48\textwidth}
        \centering
        \includegraphics[scale=.7]{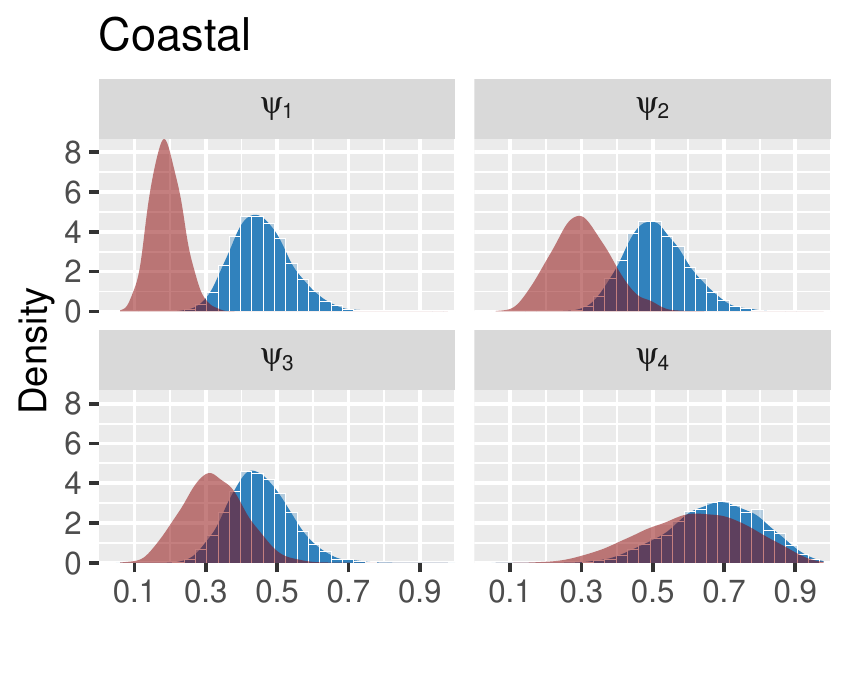}
    \end{subfigure}
    \hfill
    \begin{subfigure}[b]{.48\textwidth}
        \centering
        \includegraphics[scale=.7]{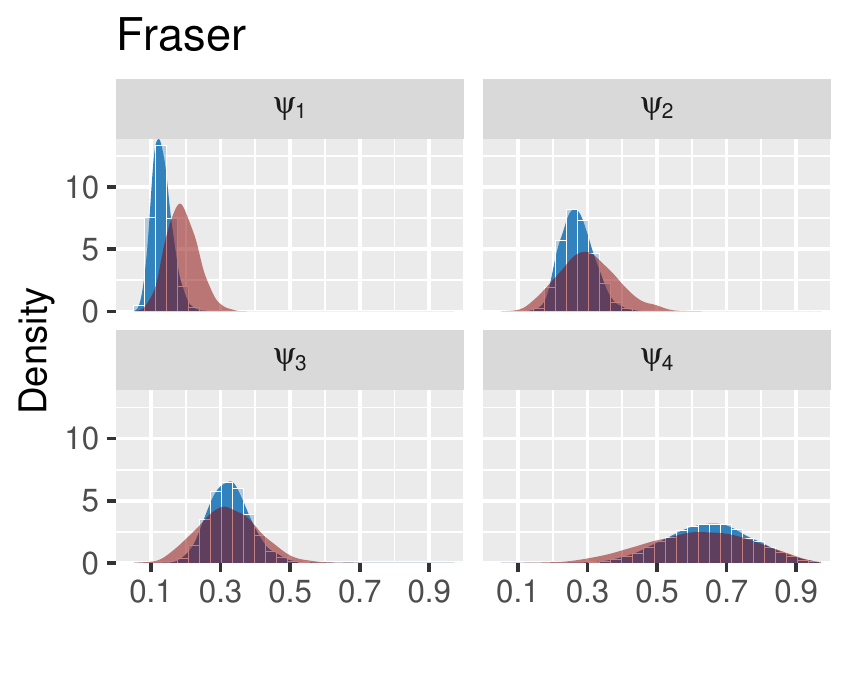}
    \end{subfigure}\\
    \begin{subfigure}[b]{.48\textwidth}
        \centering
        \includegraphics[scale=.7]{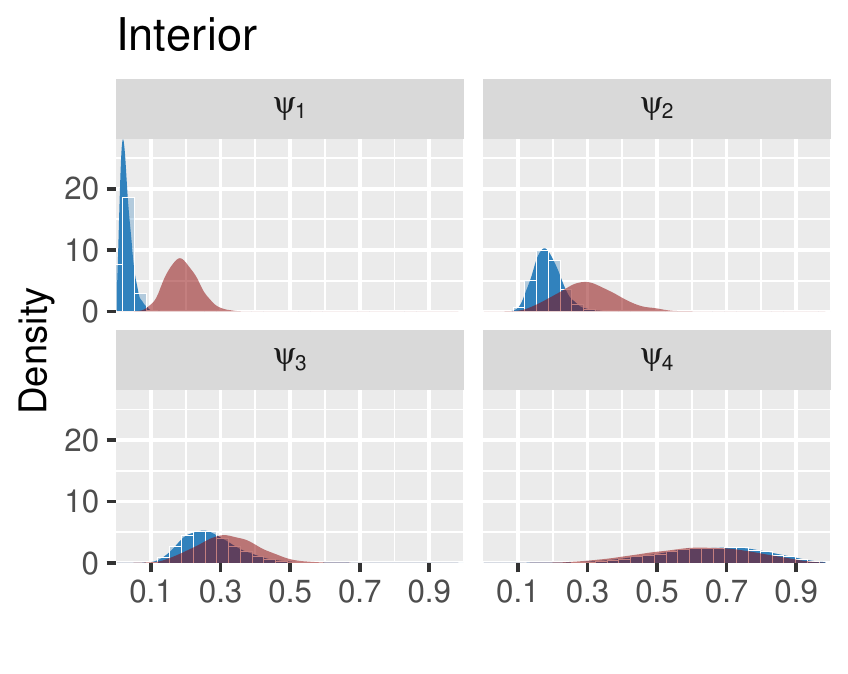}
    \end{subfigure}
    \hfill
    \begin{subfigure}[b]{.48\textwidth}
        \centering
        \includegraphics[scale=.7]{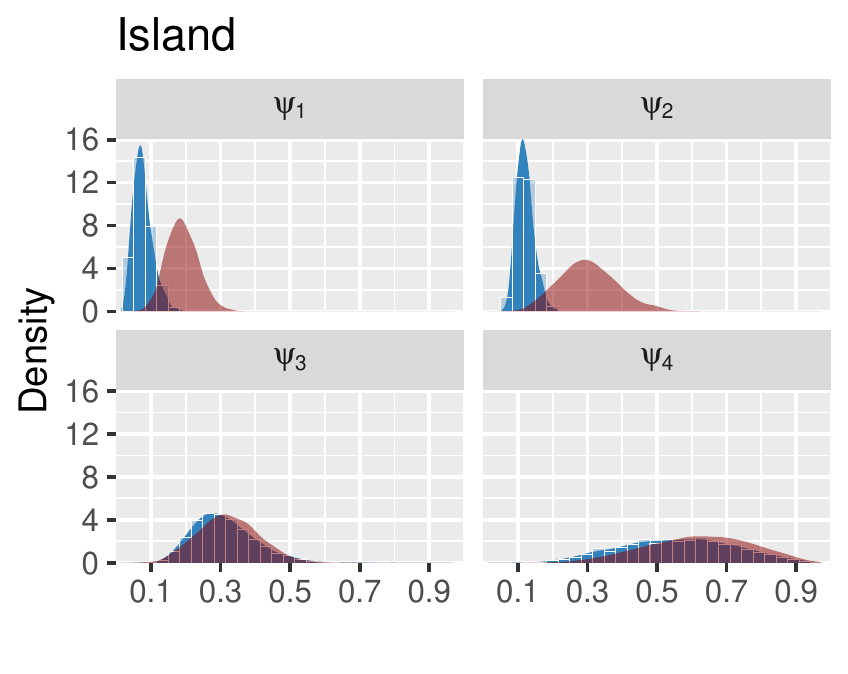}
    \end{subfigure}\\
    \begin{subfigure}[b]{.48\textwidth}
        \centering
        \includegraphics[scale=.7]{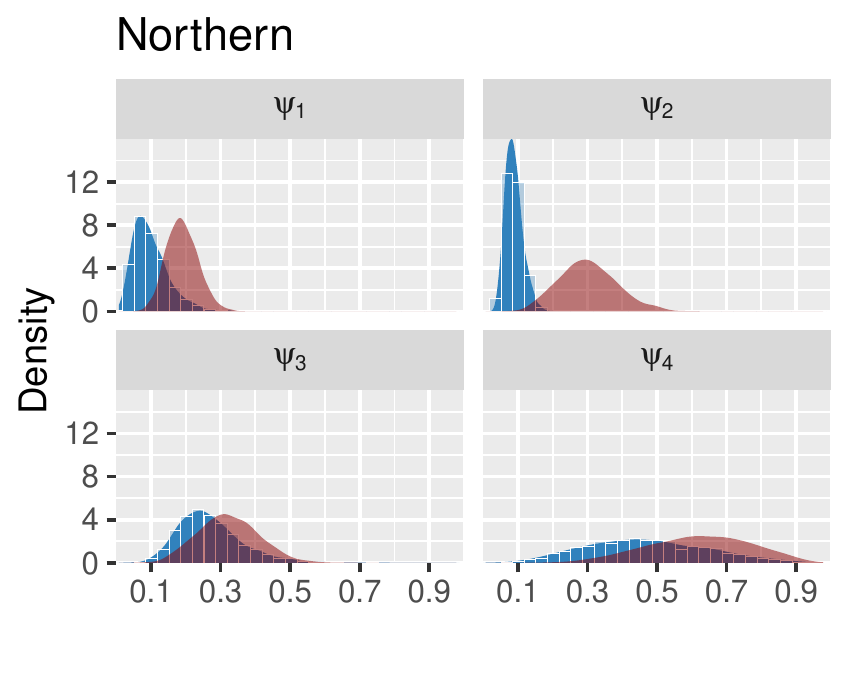}
    \end{subfigure}
    \hfill
    \vspace{-4mm}
    \caption{Comparison of estimated posterior probability densities for the proportion of anticipated cases that have been tested and reported, $\psi_1,\dots,\psi_4$ by regions (blue) and provincial-wide (red). Red and blue densities with little overlap indicate that the provincial-wide estimates are substantively different from those of the regional model.}
    \label{fig:psiComp}
\end{figure}

In this section, we compare the results between the regional and provincial-wide models, using data from March 1 to December 31 of 2020. In particular, our objective is to compare the differences in COVID-19 prevalence predicted by the regional and provincial-wide models.  We now discuss the results of this comparison depicted in Figures \ref{fig:R0b} through \ref{fig:psiComp} and in Table \ref{table:R0}.

Figure \ref{fig:R0b} displays the estimated densities for $R_{0b}$ for the hierarchical regional model in blue and the provincial-wide model in red.
There is a relatively small difference between the posterior means of these densities, which are 2.98 and 2.95
for the hierarchical regional model and the provincial-wide model, respectively. The posterior density for $R_{0b}$ in the hierarchical regional model is more concentrated around its mean, which is sensible since data from all five regions help inform its estimate. Notably, while the prior for $R_{0b}$ had its mode at 2.6 for both models, the modes of the corresponding posteriors have shifted to the 2.9--3.0 range.  This similarity in $R_{0b}$ suggests that differences in the computed $R_0$ between the models can be primarily attributed to the $f$ parameters.
 
 The plots in Figure \ref{fig:fComp} showcase the differences between the regional $f^i_2,\dots,f^i_7$ and the provincial-wide $f_2,\dots,f_7$, where we recall that $f^i_j$ represents the relative reduction of contacts in region $i$ during phase $j$.  We observe from Figure \ref{fig:fComp} that within the urban regions of Fraser and Coastal, the differences in regional and provincial-wide $f_i$ are minimal. Conversely, there are significant variations between the regional and provincial-wide $f_i$ in the more rural regions of Island, Interior and Northern.
The MCMC samples for the hierarchical regional $R_{0b}$ and the provincial-wide $R_{0b}$ (Figure \ref{fig:R0b}), together with the regional $f^i_j$ and the provincial-wide $f_j$ (Figure \ref{fig:fComp}), were used in equation \eqref{R0} to compute Table \ref{table:R0}. These are the posterior means and 95\% credible intervals of $R^i_0$, along with the regional weighted average $R_0$ and the provincial-wide $R_0$, for each period associated with the $f^i(t)$ phases (Table \ref{Phases}). Non-overlapping credible intervals indicate that there is a high probability 
 of different $R^i_0$ values for that phase. Such differences can be seen when comparing the provincial-wide and regional models (e.g., Interior and Northern differ from BC-wide $R_0$ for the $f_7$ phase), and also when comparing different regions (e.g., Interior and Island have different $R_0$'s for the $f_7$ phase) within the regional model. 
 There is a reasonable agreement between the regional weighted average $R_0$ and the provincial-wide $R_0$, except for possibly phase $f_3$. 
Given the proportionality of $R_0^i$ to $f^i(t)$ in equation \eqref{Just:R0}, the substantial increase in $f_3$ results in a large increase in the regionalized basic reproduction number, $R_0^i$, within the Interior region. This increase is reflected in Table \ref{table:R0}, resulting in a statistically significant difference from the provincially estimated $R_0$. This increase can be attributed to the large increase in cases occurring in the Interior region in early July. We provide further discussion of the results related to the Interior region at the end of this section.

\begin{table}[h!]
\centering
\footnotesize
\begin{tabular}{|c|c|c|c|c|c|c|}
\hline
 Region & $f_2$ & $f_3$ & $f_4$ & $f_5$ & $f_6$ &  $f_7$ \\
\hline
\hline
Coastal $R_0$  &    0.46 & 1.47 & 1.32 & 0.78 & 1.73 & 0.78 \\
 & (0.39,0.54) & (1.35,1.61) & (1.28,1.36) & (0.7,0.86) & (1.58,1.88) & (0.72,0.84) \\
\hline
Fraser $R_0$  &   0.65 & 1.11 & 1.21 & 1.22 & 1.59 & 0.84 \\
 & (0.59,0.71) & (1.04,1.19) & (1.18,1.25) & (1.15,1.30) & (1.48,1.71) & (0.79,0.90) \\
\hline
Interior $R_0$  &    0.30 & 2.21 & 0.92 & 1.36 & 1.75 & 1.11 \\
& (0.23,0.39) & (1.96,2.44) & (0.87,0.98) & (1.17,1.55) & (1.51,2.01) & (0.99,1.25) \\
\hline
Island $R_0$  &   0.23 & 1.64 & 1.18 & 0.86 & 2.27 & 0.80 \\
 & (0.15,0.35) & (1.31,2.00) & (1.08,1.28) & (0.71,1.04) & (2.01,2.48) & (0.72,0.89) \\
\hline
Northern $R_0$  &    0.46 & 1.31 & 1.34 & 0.64 & 2.01 & 1.18 \\
 & (0.33,0.61) & (1.05,1.60) & (1.26,1.42) & (0.52,0.78) & (1.75,2.27) & (1.07,1.29) \\
\hline
\hline
Regional Weighted   & 0.47 & 1.47 & 1.20 & 1.04 & 1.79 & 0.88 \\
Average $R_0$ & (0.42,0.52) & (1.39,1.56) & (1.17,1.22) & (0.99,1.09) & (1.71,1.87) & (0.85,0.92) \\
\hline
\hline
BC wide $R_0$  & 0.55 & 1.26 & 1.24 & 1.06 & 1.65 & 0.86 \\
 & (0.50,0.61) & (1.19,1.33) & (1.21,1.26) & (0.99,1.12) & (1.55,1.75) & (0.81,0.91) \\
\hline
\end{tabular}
\caption{\label{R0 Values} For each period corresponding to the $f_2, \ldots, f_7$ phases (columns), estimates (posterior means) of the regionalized basic reproduction number $R^i_0$ computed from the MCMC samples of $f^i_j$ and $R_{0b}$ for the regional model. The corresponding weighted average of the $R^i_0$ for the province based on the regional populations is also shown. The final row lists the estimates of the basic reproduction number $R_0$ based on the MCMC samples from the provincial-wide model. The 95\% credible intervals for each quantity are shown in parentheses.}
\label{table:R0}
\end{table}

Figure \ref{fig:prev} highlights the discrepancies between the regional and provincial-wide model prevalence of COVID-19 between March 1 and December 31 of 2020. First, we see vast differences in prevalence in the Interior region, Island region and Northern region. Within the Island region for example, the provincial-wide model grossly overestimates the number of cases from August through December 31. Whereas in the Interior and Northern regions, the provincial-wide model predicts decline through late November and December, when in fact the prevalence in these two regions were increasing.
Recalling that the $R_0$ values were significantly different between the provincial-wide and regional models during the $f_7$ phase, significant increase in prevalence were predicted in the Northern and Interior regions. To verify that the regional model accurately fits the case data, we next look at Figure \ref{fig:count}.

In Figure \ref{fig:count}, we observe that the regional model is in good agreement with the reported cases, with the vast majority of reported case counts lying within the $90\%$ posterior probability bands. Again, the Interior outbreak in early July 2020 is a notable exception. Overall, these plots further support that the regional model provides a more accurate description of the modelled prevalence as illustrated in Figure \ref{fig:prev}.  

\newpage

Finally, Figure \ref{fig:psiComp} plots the estimated densities of  $\psi_1,\dots,\psi_4$, namely the proportion of anticipated cases that have been tested and reported, shown in blue for each region. Again for comparison, the corresponding density estimates for these parameters in the provincial-wide model are shown in red.  There are some apparent differences between the regional and provincial-wide estimates for all four periods, and most noticeably for $\psi_1$ and $\psi_2$. As these are the early weeks of the pandemic, this indicates that there may be uneven testing availability in the province, resulting in posteriors for $\psi_1$ and $\psi_2$ that vary considerably from the priors assigned to them.  

Furthermore, the assumed initial conditions for the model (which were simply scaled regionally due to low initial case counts) may contribute to these differences.  Once testing became widely available (April 21 onwards, represented by $\psi_4$), it appears that there is considerable uncertainty: while $\psi_4$ has increased relative to $\psi_3$,  its plausible range of values in the posterior is quite wide, between 0.3--0.9 for most regions and also in the provincial-wide model.
 
Nonetheless, the $\psi_4$ posteriors are not purely reflecting the prior chosen (which had a mode at 0.4 from Section \ref{Stat Model}) and regional differences are still apparent; e.g., Northern has most of its $\psi_4$ posterior probability density from 0.1--0.7, while the corresponding range is 0.5-0.9 for Coastal.

Combining the results showcased in Figures \ref{fig:fComp}, \ref{fig:prev}, \ref{fig:count} and Table \ref{table:R0}, we see that the provincial-wide model is unable to adequately capture the modelled prevalence throughout all regions. In particular, since data from all regions were grouped together, smaller outbreaks, such as the July outbreak in the Interior region, did not make a significant impact on the estimated provincial $R_0$. Specifically, the estimated mean of the provincial $R_0$ during the $f_3$ phase was $1.26$, while in the Interior region, the regional model yielded an estimated $R_0^i$ mean of $2.21$, with non-overlapping $95\%$ credible intervals. Additionally, Table \ref{table:R0} indicates several other statistically significant differences between the provincial and regional models, for instance during the $f_5$ phase for the Northern and Island regions.
 
While the hierarchical regional model accurately estimated an increase in cases for the Interior region through the end of June into July of 2020, it failed to fit the isolated increase and decrease in cases appearing during the $f_4$ phase, as seen in Figure \ref{fig:count}. Since $R_0^i$ is constant during each phase, as shown in equation \eqref{Just:R0}, our proposed modelling framework is unable to predict an increase and decrease in cases within a single phase. One explanation why the outbreak was immediately followed by a rapid decrease in cases was due to interventions introduced in mid-July of 2020 affecting restaurants, bars, nightclubs, rental properties and house boats, as reported by \cite{CBCNews21a}.
If this additional intervention had been counted as a provincial  lockdown phase, our hierarchical regional model may have been able to better estimate the increase and decrease in cases during this short period.

Overall, the results presented in this section highlight that using strictly provincial modelling to estimate prevalence may miss localized trends within regions. This is illustrated by Figure \ref{fig:prev} during the $f_7$ phase in December of 2020. The provincial-wide model suggests a significant decline in cases, while the regional model suggests a drastic increase in cases within the Northern and Interior regions. Therefore, if only provincial-wide modelling is utilized, loosening of provincial restrictions may be recommended. However, as shown by the regional model, such a recommendation could be catastrophic within the Interior and Northern regions. 
\subsection{Comparison in Prevalence Prediction between Provincial-wide and Hierarchical Regional Models}

\begin{figure}[h!]
    \centering
    \begin{subfigure}[b]{0.44\textwidth}
        \centering
        \includegraphics[scale=0.7]{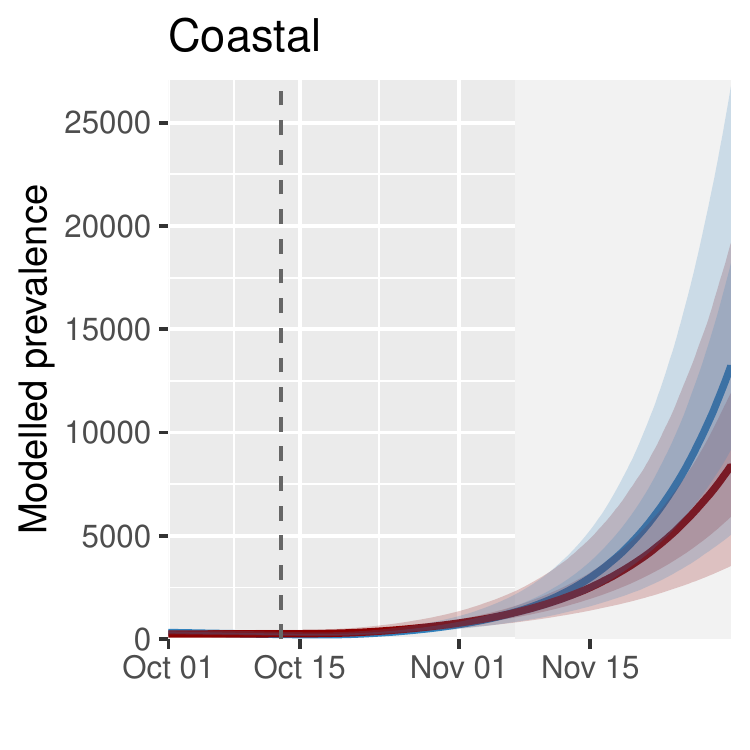}
    \end{subfigure}
    \hfill
    \begin{subfigure}[b]{0.44\textwidth}
        \centering
        \includegraphics[scale=0.7]{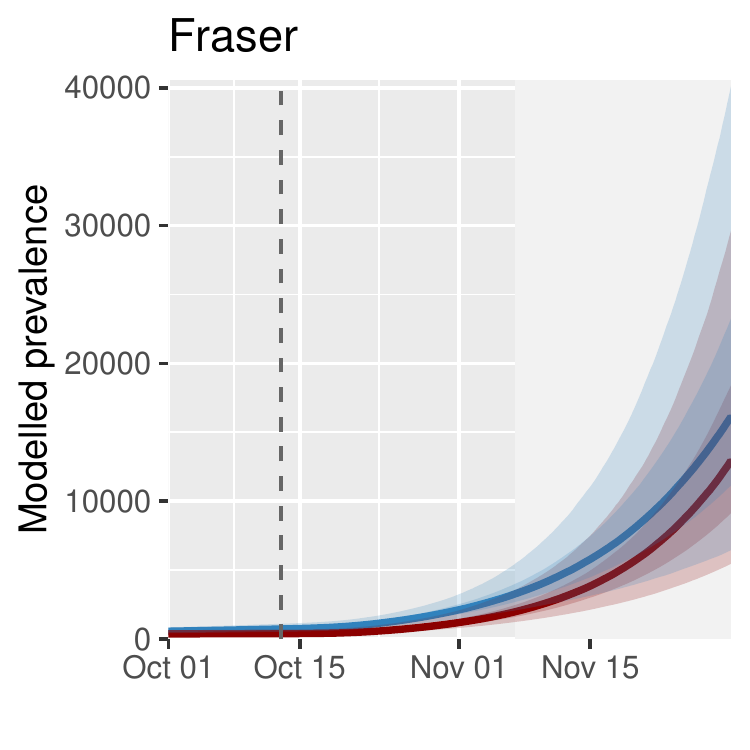}
    \end{subfigure}\\
    \begin{subfigure}[b]{0.44\textwidth}
        \centering
        \includegraphics[scale=0.7]{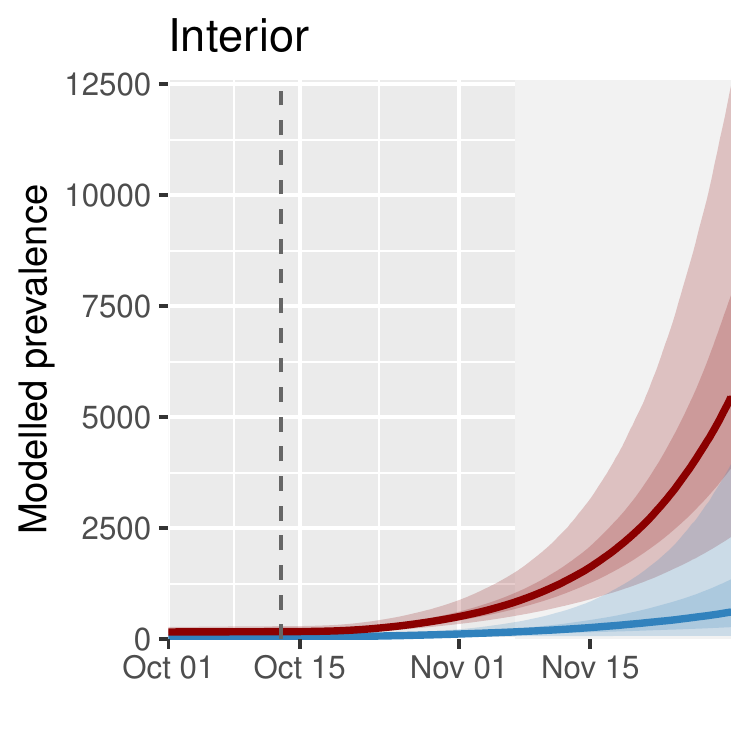}
    \end{subfigure}
    \hfill
    \begin{subfigure}[b]{0.44\textwidth}
        \centering
        \includegraphics[scale=0.7]{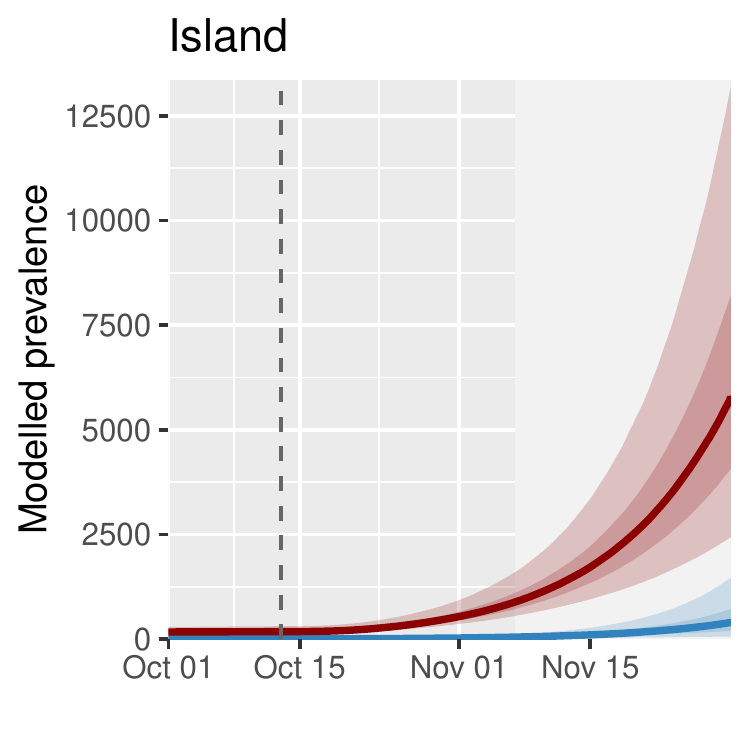}
    \end{subfigure}\\
    \begin{subfigure}[b]{0.44\textwidth}
        \centering
        \includegraphics[scale=0.7]{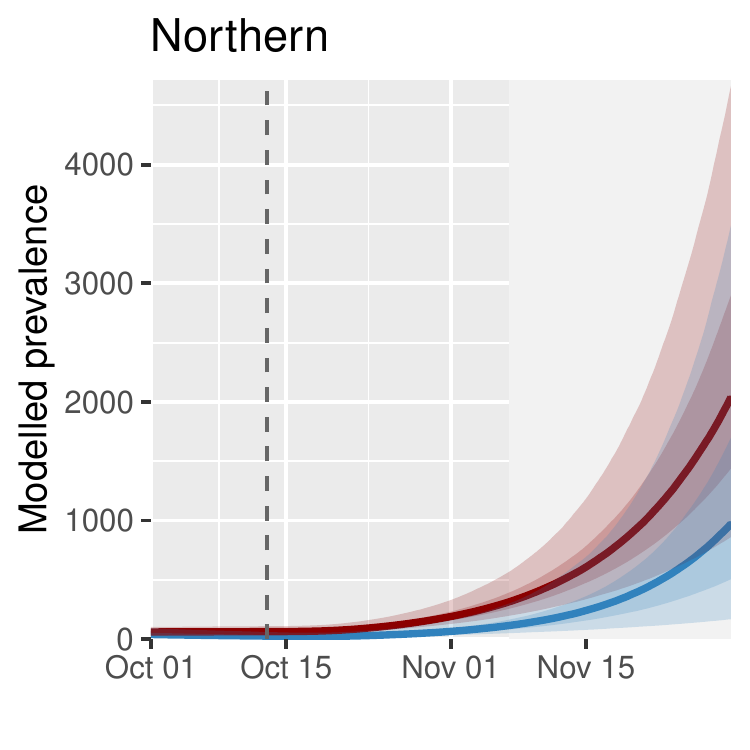}
    \end{subfigure}\hfill
    \vspace{-4mm}
    \caption{Comparison of predicted modelled prevalence by regions (blue) versus provincial-wide (red). Dotted line indicate start of $f_6^i$ on October 12th and the light grey regions indicate prediction region starting from November 7th. Again, solid lines indicate the respective mean and shaded regions indicate respective ranges with 50\% and 90\% posterior probability.}
    \label{fig:pred}
\end{figure}

For this comparison, the two models were fitted using data from March 1 to November 7 of 2020 and subsequently used to predict the prevalence of COVID-19 cases in the time period between November 8 and November 30 of 2020. The resulting plots are shown in Figure \ref{fig:pred}. In many ways, these plots are the most impactful of our results, as we now explain.

The epidemiological model described by \eqref{Model} and \eqref{SDModel}, with parameters estimated via the statistical methodology presented in Section \ref{Stat Model}, yield our best estimates of how the pandemic is evolving at a given time, along with credible intervals quantifying the associated uncertainty with those estimates. We verified that our modeling approach allows us to fit the regional data well, as discussed in the previous subsection. 
The final step is to illustrate the utility of our model in making predictions. The predicted time period is shown in the light grey portions of Figure \ref{fig:pred}. 

Focusing our attention to the rural regions (Interior, Island and Northern) in Figure \ref{fig:pred}, we see an overestimation of cases predicted by the provincial-wide model. Recall from Section \ref{sec:modelInput}, the provincial-wide model groups together all cases within the province, predicts the province-wide prevalence, and then scales the cases to the individual regions based on their population ratio. Since the cases per capita in the urban regions (Fraser and Coastal) are the highest in the province, scaling the provincial results to the individual regions leads to an overestimation and an underestimation in the rural and urban regions, respectively. From the results of Figure \ref{fig:pred}, the provincial-wide model suggests a drastic increase in cases throughout the province if additional mitigation measures are not implemented. In contrast, the regional model provides different prevalence predictions in certain rural regions. Specifically, we see no overlap with the 90\% posterior band for the Island region and minor overlap with the 90\% posterior band for the Interior region.
Overall, both the provincial-wide and regional models suggest that all regions should introduce additional mitigation measures. However, the regional model suggests further mitigation efforts should take place in the urban regions\footnote{It may also be possible for the provincial-wide model to improve its prediction, provided regional data is incorporated.}. These efforts could include an increase in testing, additional contact tracing and travel restrictions in the Fraser and Coastal regions to prevent the further spread of COVID-19.

As discussed in the introduction section of the paper, migration between regions has been shown to be a significant contributor to the spread of COVID-19. Therefore, in order to protect the more rural regions from further increase in cases, a reduction in migration from Coastal and Fraser regions would be advisable. In the absence of such travel restrictions, the rural regions are vulnerable to a sudden increase in cases due to an influx of infectious individuals from the Coastal or Fraser regions. Moreover, we observe from Figure \ref{fig:prev} that the rise in cases throughout November was first observed in the Fraser and Coastal regions. This is then followed by increases in the Northern region and later in the Interior and Island Regions. Hence, through regionalized predictive modelling, specific region mitigation measures can made to help prevent further spread of COVID-19.

\subsection{Concluding Remarks}

In this paper, we adapted the provincial-wide model of \cite{AndersonAE20} to a regional model and studied their differences. In contrast to \cite{AndersonAE20}'s provincial-wide model, we introduced an additional hierarchical structure to facilitate regional modelling, with specific regional parameters.  Our results showed that the proposed hierarchical model can effectively fit the regional case data and highlighted important differences in both prevalence estimates and prediction versus the provincial-wide model. For example, when comparing the provincial and regional models, statistically significant differences in the $R_0^i$ estimates were observed in rural regions during certain lockdown phases. Our results also indicate that the regional model was able to detect smaller trends overlooked by the provincial-wide model, such as the rise in cases within the Interior region throughout late June and early July of 2020. Furthermore, the provincial-wide model was shown to overestimate prevalence in certain rural regions and underestimate prevalence in the urban regions of BC. This suggests that regionalized models can provide valuable insight potentially overlooked by using a provincial-wide model and help guide future interventions to reduce the spread of COVID-19 in BC. The value of regional modelling of COVID-19 has also been demonstrated in other recent work, such as \cite{KaratayevAE20} and \cite{BakhtaAE21}.
Moreover, the application of our proposed regional model is not restricted to only British Columbia. Upon modifying parameters to specific regions, our proposed regional model can be applied to other parts of Canada or other countries, to enhance regional predictions of COVID-19 prevalence. A public Github repository with the R code is provided to facilitate such usage, as given in the Code Availability statement at the end of the paper.

Nonetheless, our proposed model has certain limitations. The ODE systems do not account explicitly for migration between regions. Such migration could have played a key role especially during the initial spread of COVID-19 to the different regions of the province.  Also, we did not attempt to incorporate time-varying or region-specific parameters $u_d$ and $u_r$.  For example, $u_d$ and $u_r$ may have changed as the pandemic progressed due to lockdown fatigue.  We also note that data on the number of tests performed and positivity rates have not been used in our model. Intuitively, such data may have some relationship to $\psi_i$, but their actual role is difficult to determine without random testing. Overall, access to additional data could greatly improve the accuracy of the proposed model. For example, the parameter $f^i(t)$ currently accounts for all the mitigation measures that limit the spread of COVID-19, such as physical distancing, masks, hand washing and contact tracing. Although these mitigation measures remained fairly constant during our period of study, they still had some influence on the resulting estimates for $f^i(t)$. Therefore, access to specific data on the other mitigation measures could allow us to better determine the degree to which the parameter $f^i(t)$ can be attributed to physical distancing. As discussed in Section \ref{sec:prevalence}, our model only accounts for the official lockdown phases imposed provincially between March 1 and December 31 of 2020. However, other interventions, such as changes in protocols for bars and restaurants, may have impacted the provincial and regional case counts during these months. Therefore, our regional model could be improved by accounting for all such interventions.
Finally, it is important to note that not all outbreaks are a result of an increase in contacts. Certain outbreaks, known as superspreaders, emerge from parties or gatherings, where significantly more infections occur than anticipated. Recent models account for superspreaders as highly infectious individuals as discussed in \cite{KaratayevAE20}, or individuals with a high number of contacts as studied by \cite{ChenAE20}. Additionally, superspreader events have been studied using an SEIR model in conjunction with cellular network data by \cite{ChangAE2021}. In contrast, our ODE model is unable to account for superspreader events, or superspreading individuals, and therefore may misrepresent  outbreaks resulting from superspreaders as a general increase in contacts. 

Furthermore, upon finishing our work, we have noticed that the BC Centre for Disease Control has started reporting regional predictions in the weekly report of \cite{BCCDC21c} as of the week of April 14 of 2021, and as well as in their recent annual report of \cite{BCCDC21d}. Thus, another extension to this work would be to modify our hierarchical Bayesian epidemiological model to incorporate vaccinated individuals and additional modifications to account for more infectious variants of COVID-19. With additional change points to account for recent restrictions in 2021, the modified model would allow predictions to be made with the current 2021 data and be compared with the regional predictions from the BC Centre for Disease Control.

\section*{Code Availability}
The R code and programs that support the results of this analysis are publicly available in the Github repository at \url{https://github.com/wongswk/hierarchical-regional-covid}.

\section*{Acknowledgements}

The authors would like to thank the anonymous reviewers for their helpful comments and valuable suggestions at improving this paper.

\section*{Funding Acknowledgements}

Samuel W.K. Wong's research was partially supported by a Discovery Grant from the Natural Sciences and Engineering Research Council (NSERC) of Canada. Jennifer Tippett was a former University of Northern British Columbia (UNBC) Masters student and her research in this work were supported by the NSERC COVID-19 Supplement. Geoffrey McGregor is a UNBC postdoctoral fellow, and his research and Andy Wan's research were partially supported by the NSERC Discovery Grant Program.

\section*{Author Information and Contribution}

The authors have no competing interests declared.\vspace{4mm}\\
\textbf{Geoffrey McGregor} (UNBC)\\
\underline{Contribution}: Model Analysis, Validation, Methodology, Writing, Review and Editing\vspace{3mm}\\
\textbf{Jennifer Tippett} (UNBC)\\
\underline{Contribution}: Data curation, Writing, Review and Editing\vspace{3mm}\\
\textbf{Andy T.S. Wan} (UNBC)\\
\underline{Contribution}: Conceptualization, Validation, Methodology, Visualization, Writing, Review, Editing and Supervision\vspace{3mm}\\
\textbf{Mengxiao Wang} (University of Waterloo)\\
\underline{Contribution}: Data curation, Statistical Analysis, Visualization, Writing, Review and Editing\vspace{3mm}\\
\textbf{Samuel Wong} (University of Waterloo)\\
\underline{Contribution}: Validation, Methodology, Statistical Analysis, Visualization, Writing, Review, Editing and Supervision

\bibliographystyle{apalike}
\bibliography{refs}

\newpage
\begin{appendices}

\section{MCMC histograms for simulation study} \label{sec:simstudyplot}
\begin{figure}[!htbp]
    \centering
    \begin{subfigure}[b]{\textwidth}
        \centering
        \includegraphics[scale=0.65]{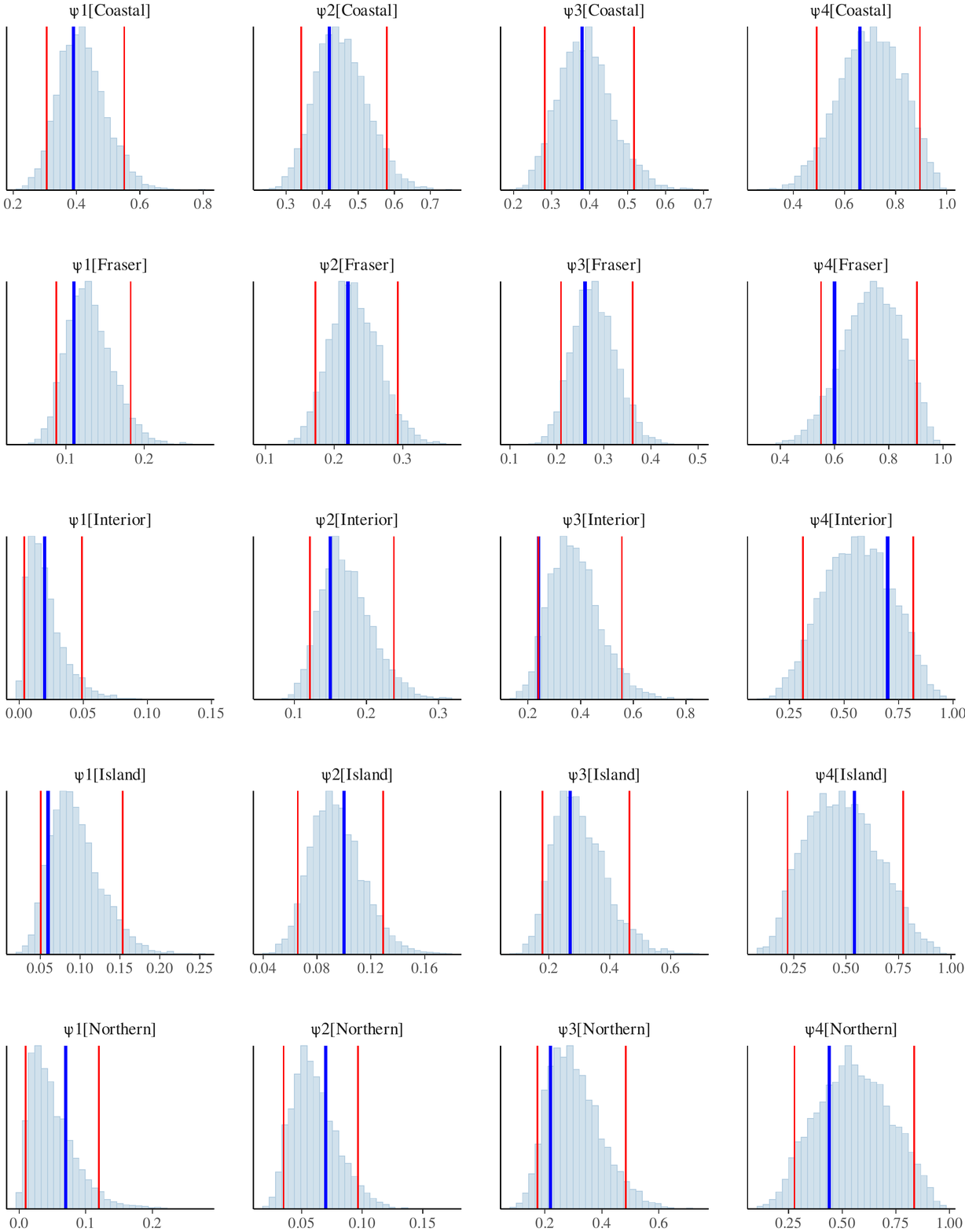}
    \end{subfigure}
    \caption{Histograms of MCMC samples for $\psi_1,\dots,\psi_4$ in the simulation study. The true value of each parameter is shown with the blue vertical line, and the bounds of the central 90\% credible interval for each parameter is shown with the red vertical lines.}
    \label{fig:simhist-regional-a}
\end{figure}
\newpage

\begin{figure}[h!]
    \centering
    \begin{subfigure}[b]{\textwidth}
        \centering
        \includegraphics[scale=0.7]{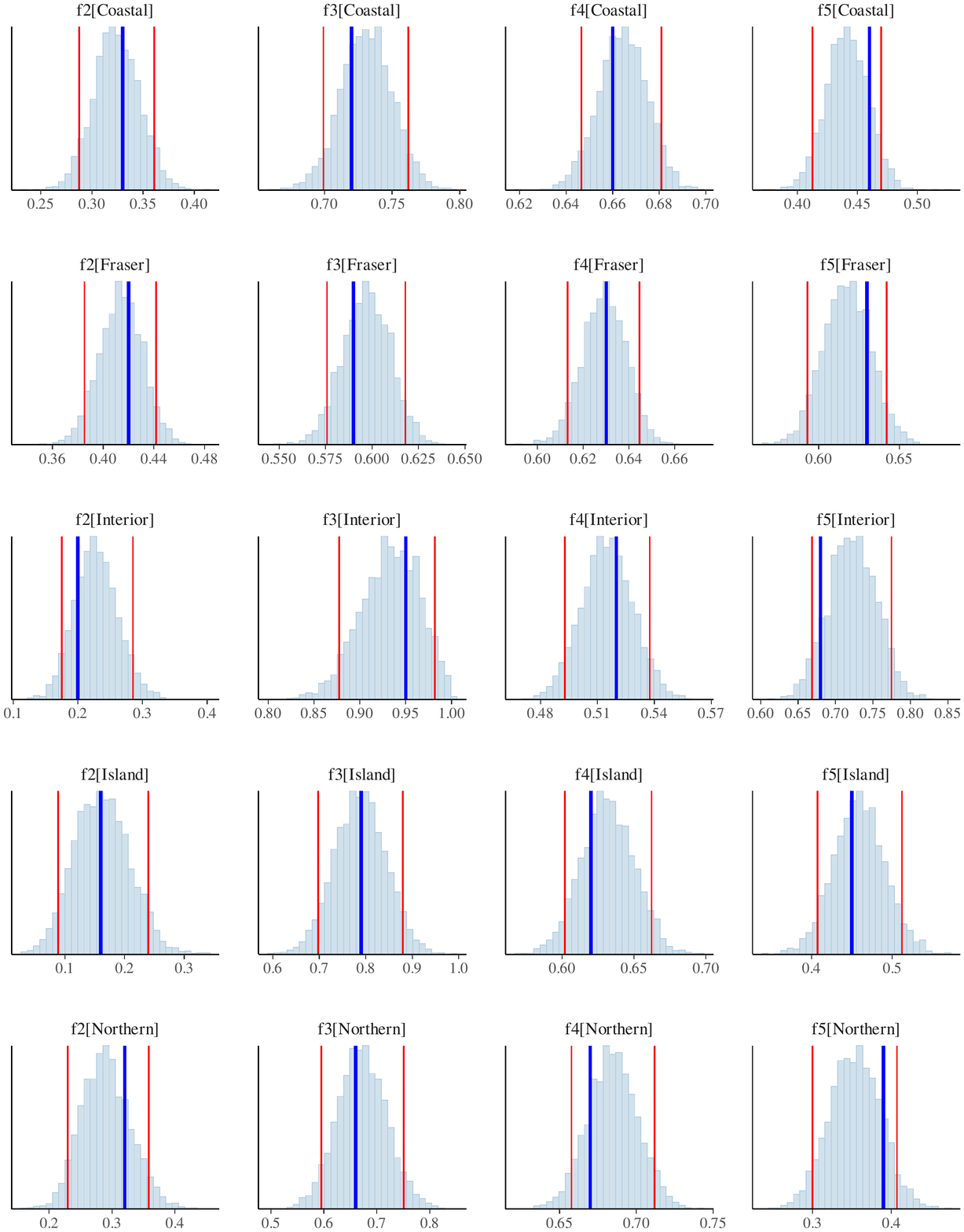}
    \end{subfigure}
    \caption{Histograms of MCMC samples for  $f_2,\dots,f_5$ in the simulation study. The true value of each parameter is shown with the blue vertical line, and the bounds of the central 90\% credible interval for each parameter is shown with the red vertical lines.}
    \label{fig:simhist-regional-b}
\end{figure}
\newpage

\begin{figure}[h!]
    \centering
    \begin{subfigure}[b]{\textwidth}
        \centering
        \includegraphics[scale=0.7]{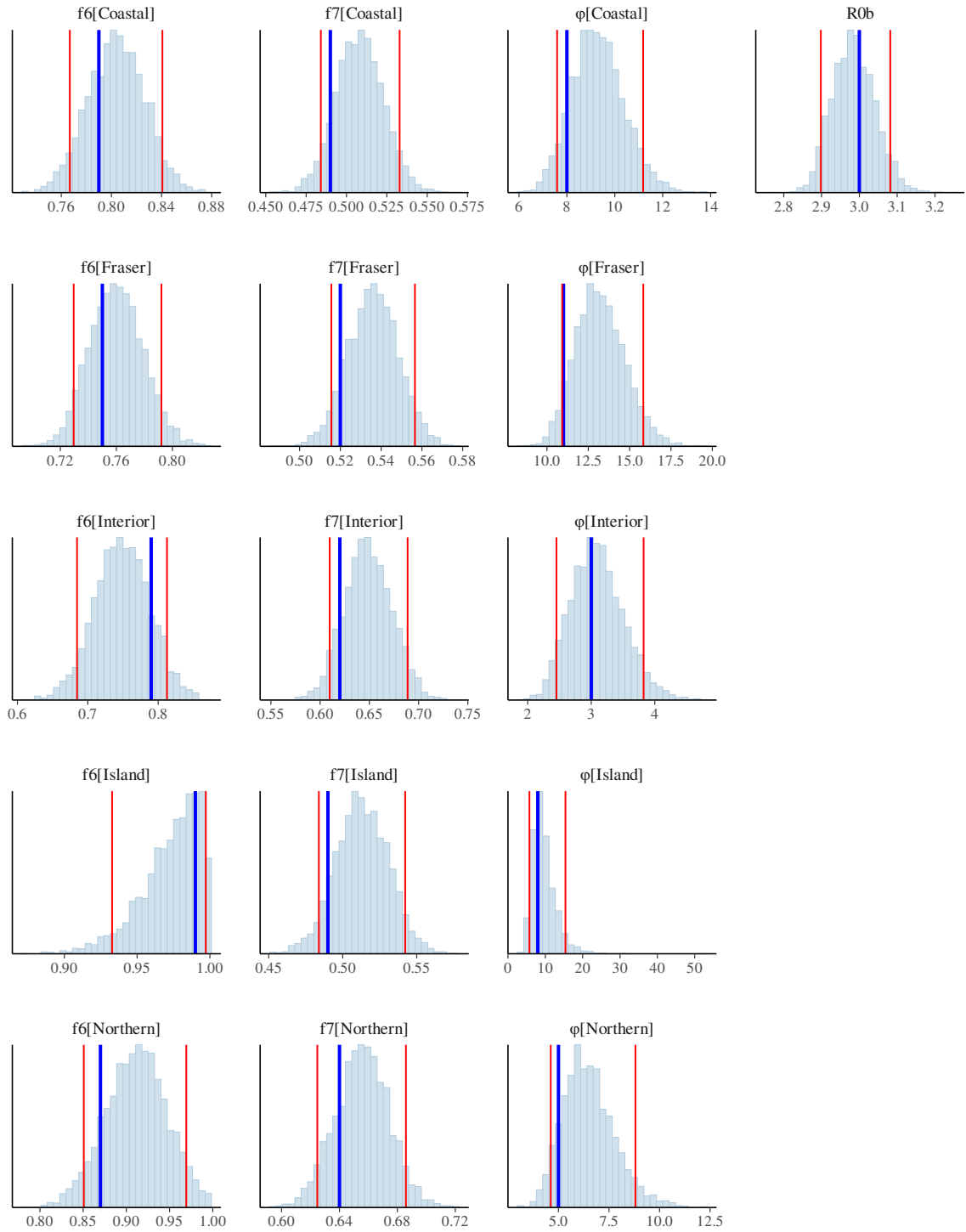}
    \end{subfigure}
    \caption{Histograms of MCMC samples for $f_6,f_7,\phi$ and hierarchical regional $R_{0b}$ in the simulation study. The true value of each parameter is shown with the blue vertical line, and the bounds of the central 90\% credible interval for each parameter is shown with the red vertical lines.}
    \label{fig:simhist-regional-c}
\end{figure}
\newpage

\section{Non-Hierarchical versus Hierarchical regional model fits comparison}
\label{sec:NHregional}
\begin{figure}[h!]
    \centering
    \begin{subfigure}[b]{\textwidth}
        \centering
        \includegraphics[scale=0.64]{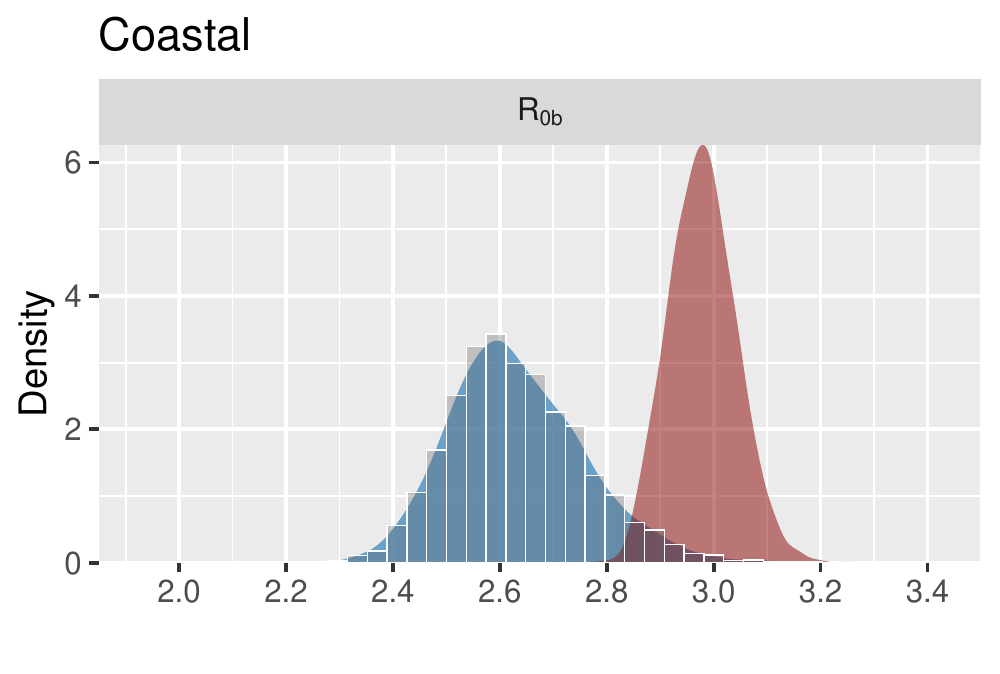}
    \end{subfigure}\\
    \begin{subfigure}[b]{\textwidth}
        \centering
        \includegraphics[scale=0.64]{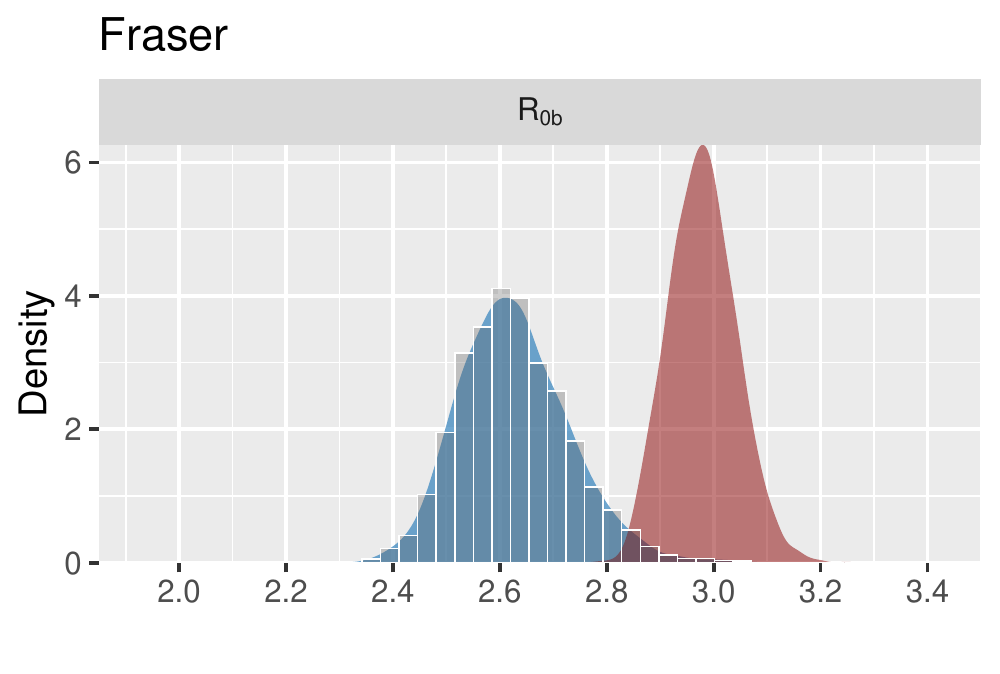}
    \end{subfigure}\\
    \begin{subfigure}[b]{\textwidth}
        \centering
        \includegraphics[scale=0.64]{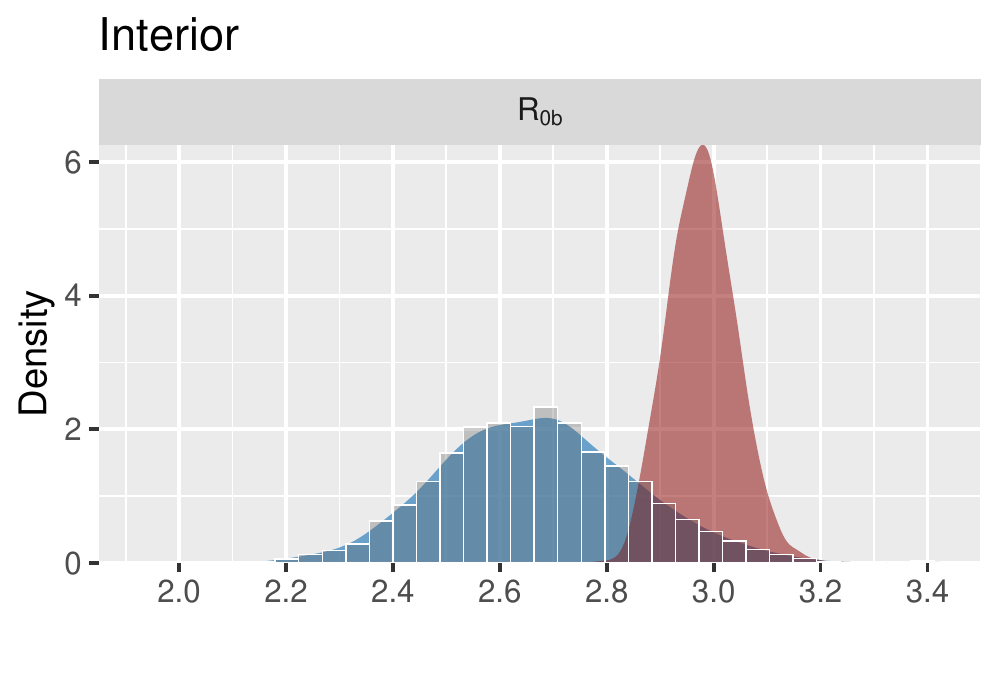}
    \end{subfigure}\\
    \begin{subfigure}[b]{\textwidth}
        \centering
        \includegraphics[scale=0.64]{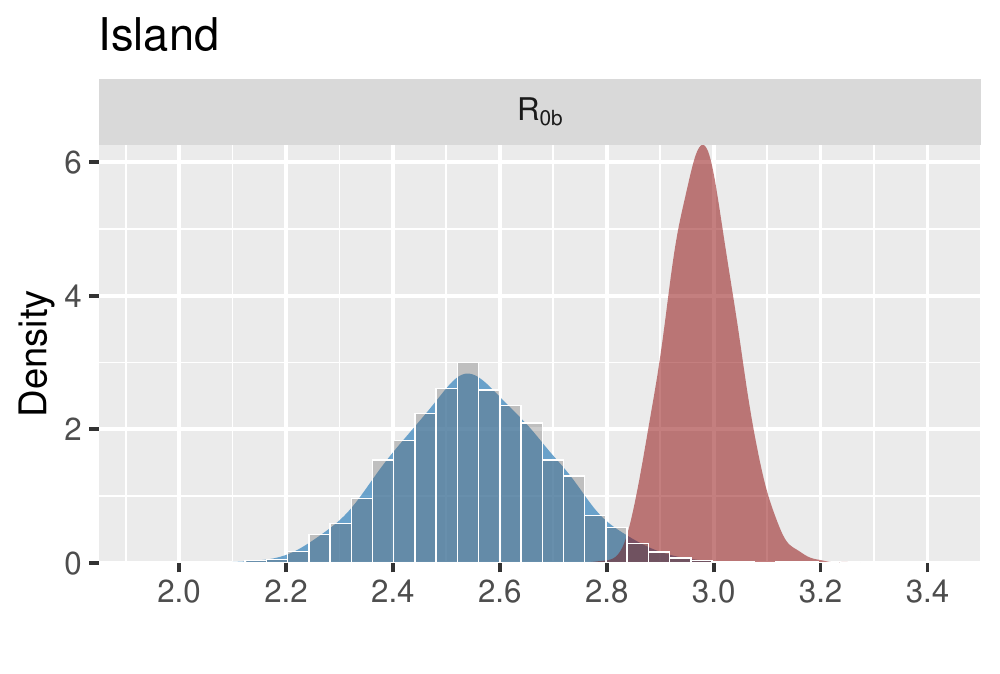}
    \end{subfigure}\\
    \begin{subfigure}[b]{\textwidth}
        \centering
        \includegraphics[scale=0.64]{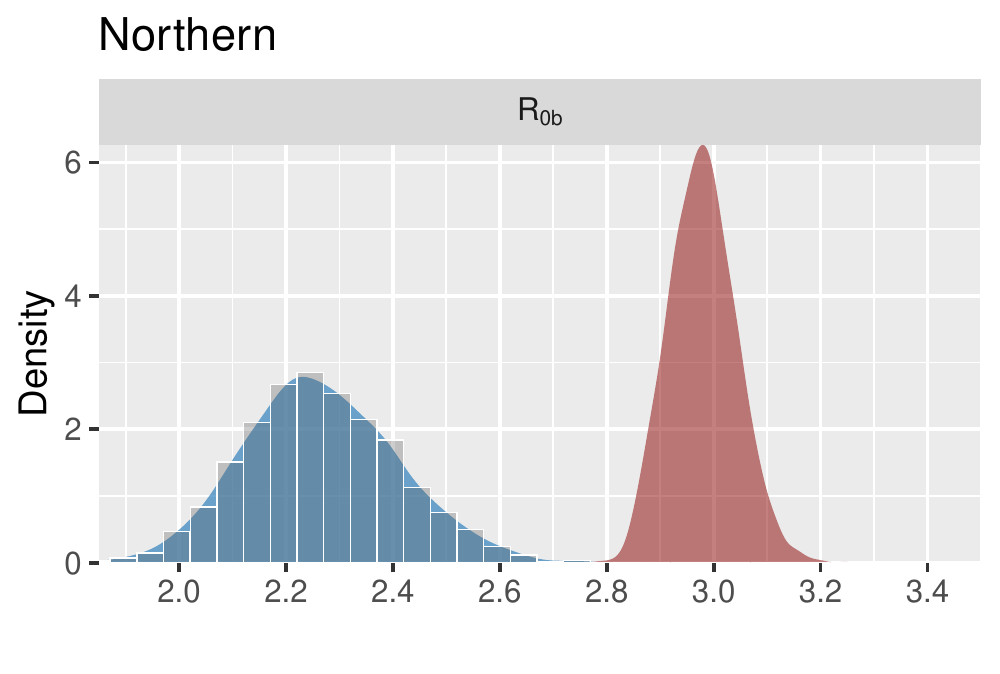}
    \end{subfigure}
    \vspace{-10mm}
    \caption{For models fitted to individual regions only, the estimated posterior density of $R_{0b}$ in each region is shown in blue, with the corresponding posterior density of $R_{0b}$ from the hierarchical regional model shown in red.}
    \label{fig:R0bCompIndiv}
\end{figure}

\newpage
\begin{figure}[h!]
    \centering
    \begin{subfigure}[b]{\textwidth}
        \centering
        \includegraphics[scale=0.68]{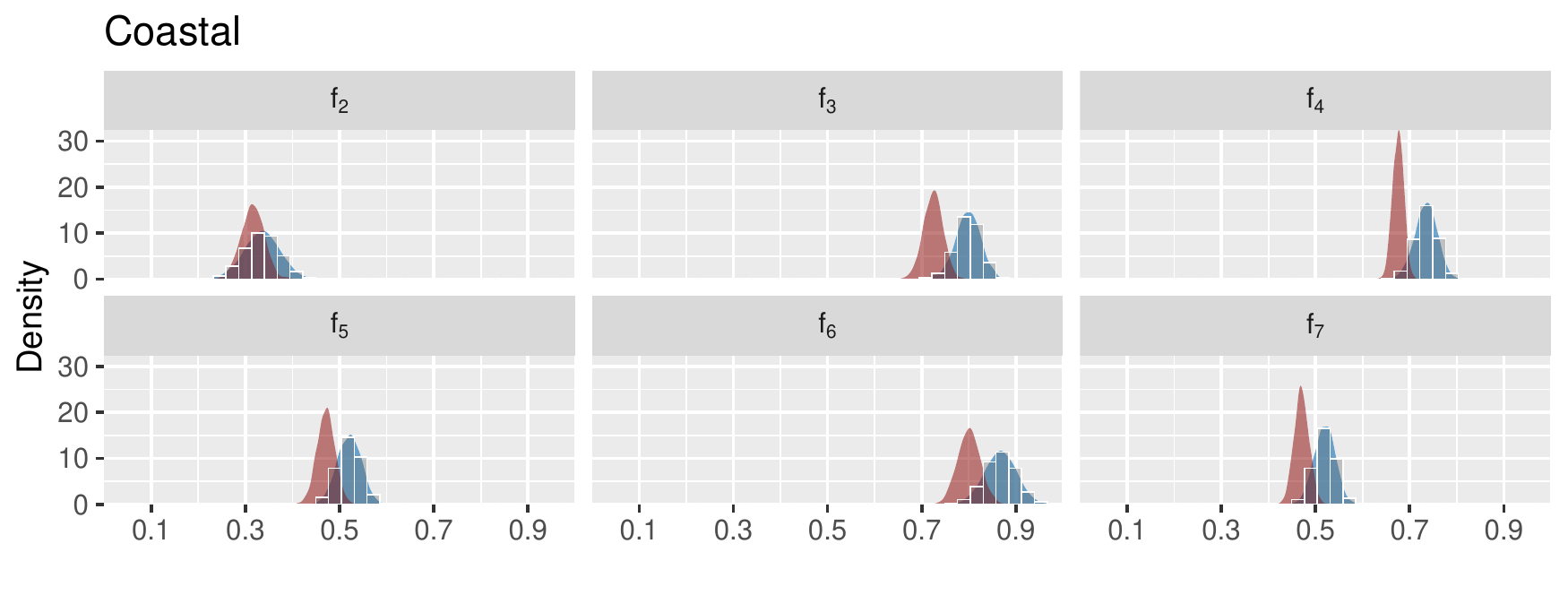}
    \end{subfigure}\\
    \begin{subfigure}[b]{\textwidth}
        \centering
        \includegraphics[scale=0.68]{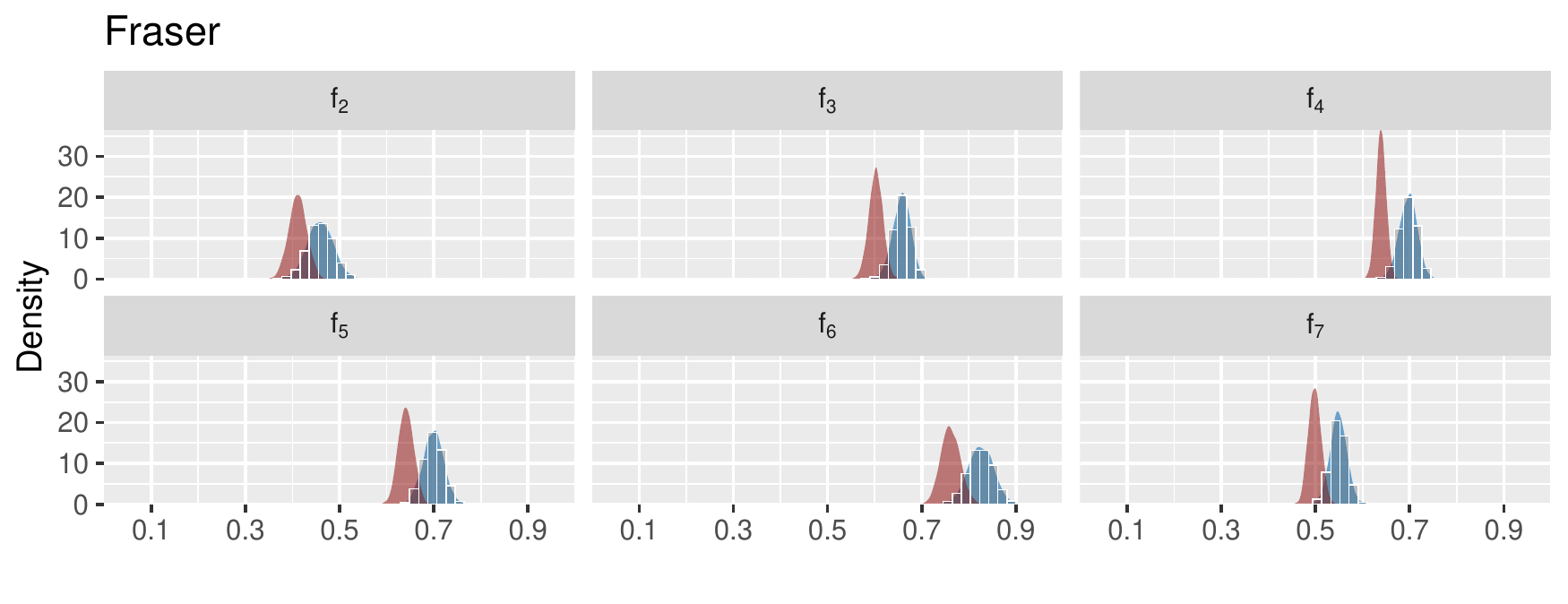}
    \end{subfigure}\\
    \begin{subfigure}[b]{\textwidth}
        \centering
        \includegraphics[scale=0.68]{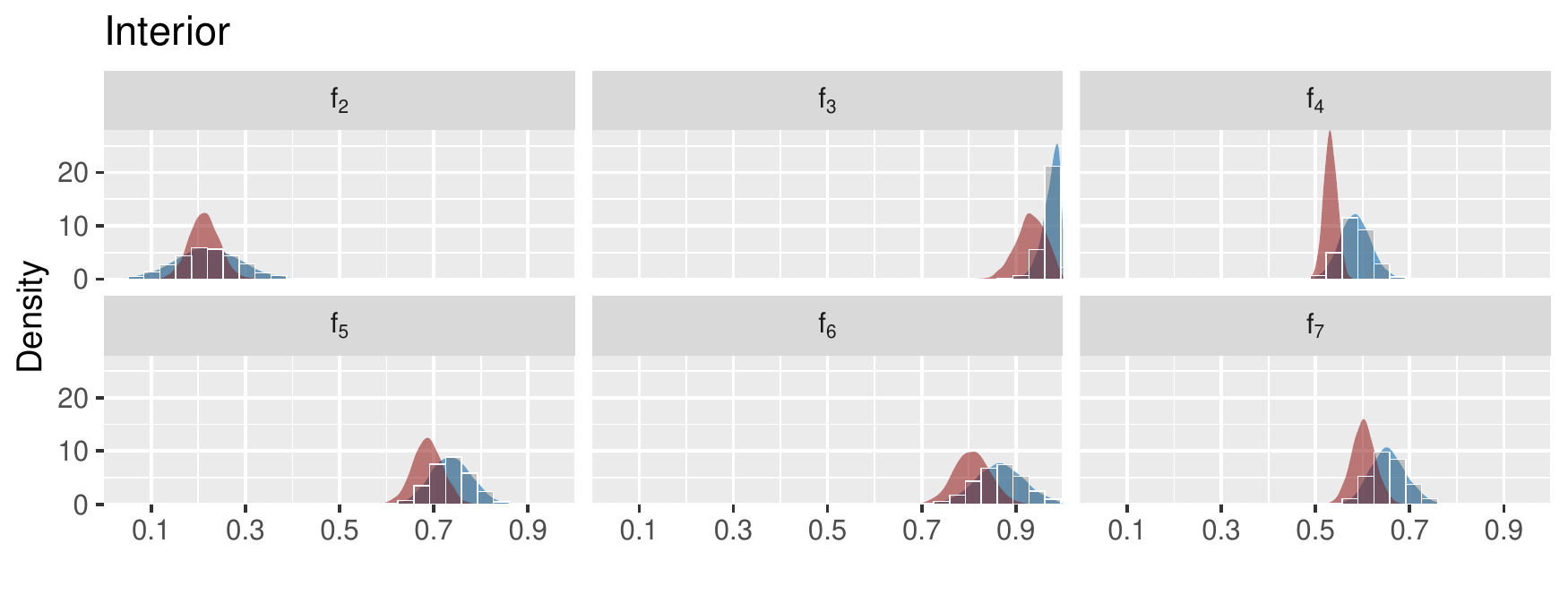}
    \end{subfigure}\\
    \begin{subfigure}[b]{\textwidth}
        \centering
        \includegraphics[scale=0.68]{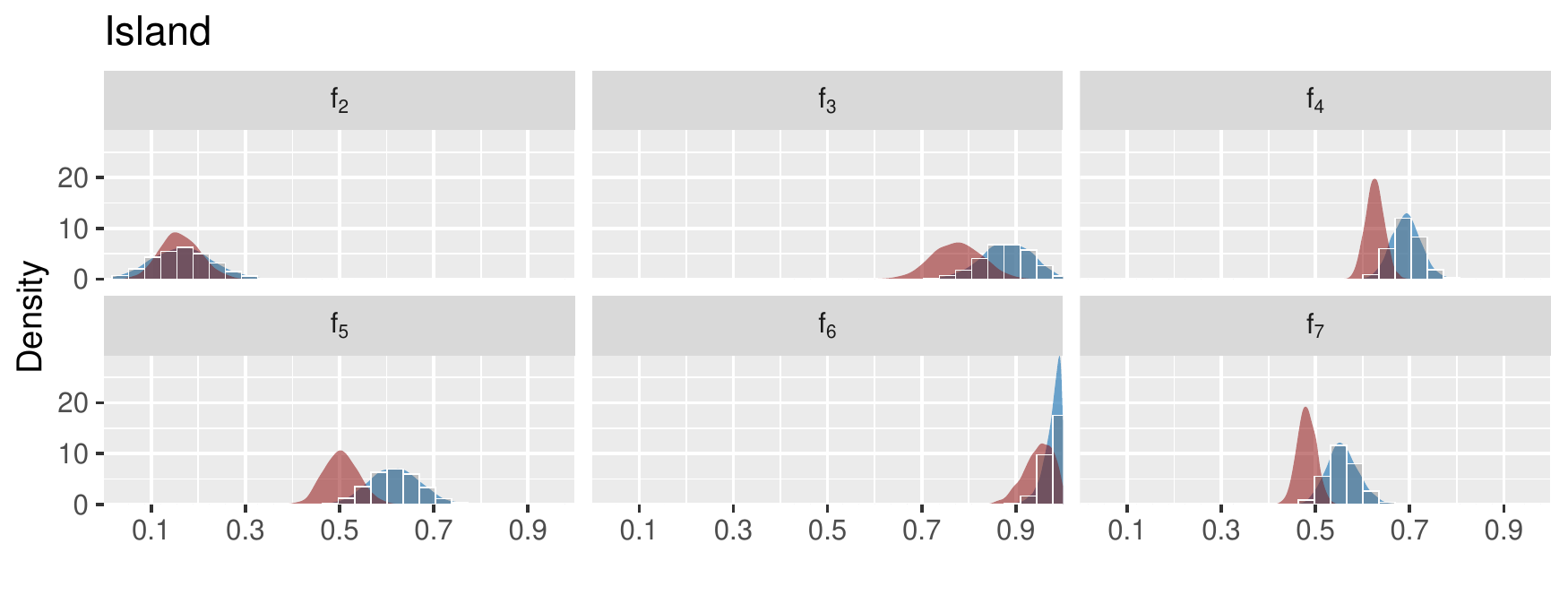}
    \end{subfigure}\\
    \begin{subfigure}[b]{\textwidth}
        \centering
        \includegraphics[scale=0.68]{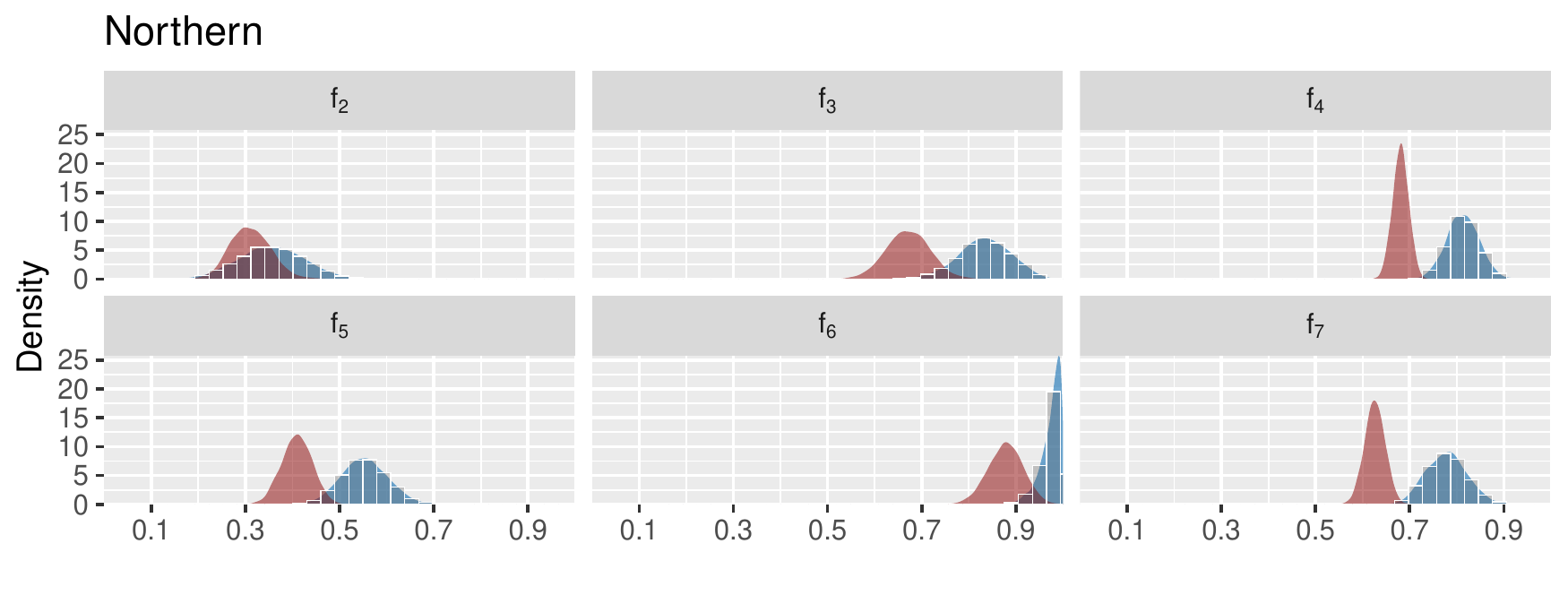}
    \end{subfigure}
    \vspace{-10mm}
    \caption{For models fitted to individual regions only, the estimated posterior densities of $f_2,\dots,f_7$ in each region are shown in blue. For comparison, the posterior densities of the same parameters from the hierarchical regional model are shown in red.}
    \label{fig:fCompIndiv}
\end{figure}

\section{Trace plots}
\label{sec:traceplot}

\begin{figure}[h!]
    \centering
    \begin{subfigure}[b]{\textwidth}
        \centering
        \includegraphics[scale=0.7]{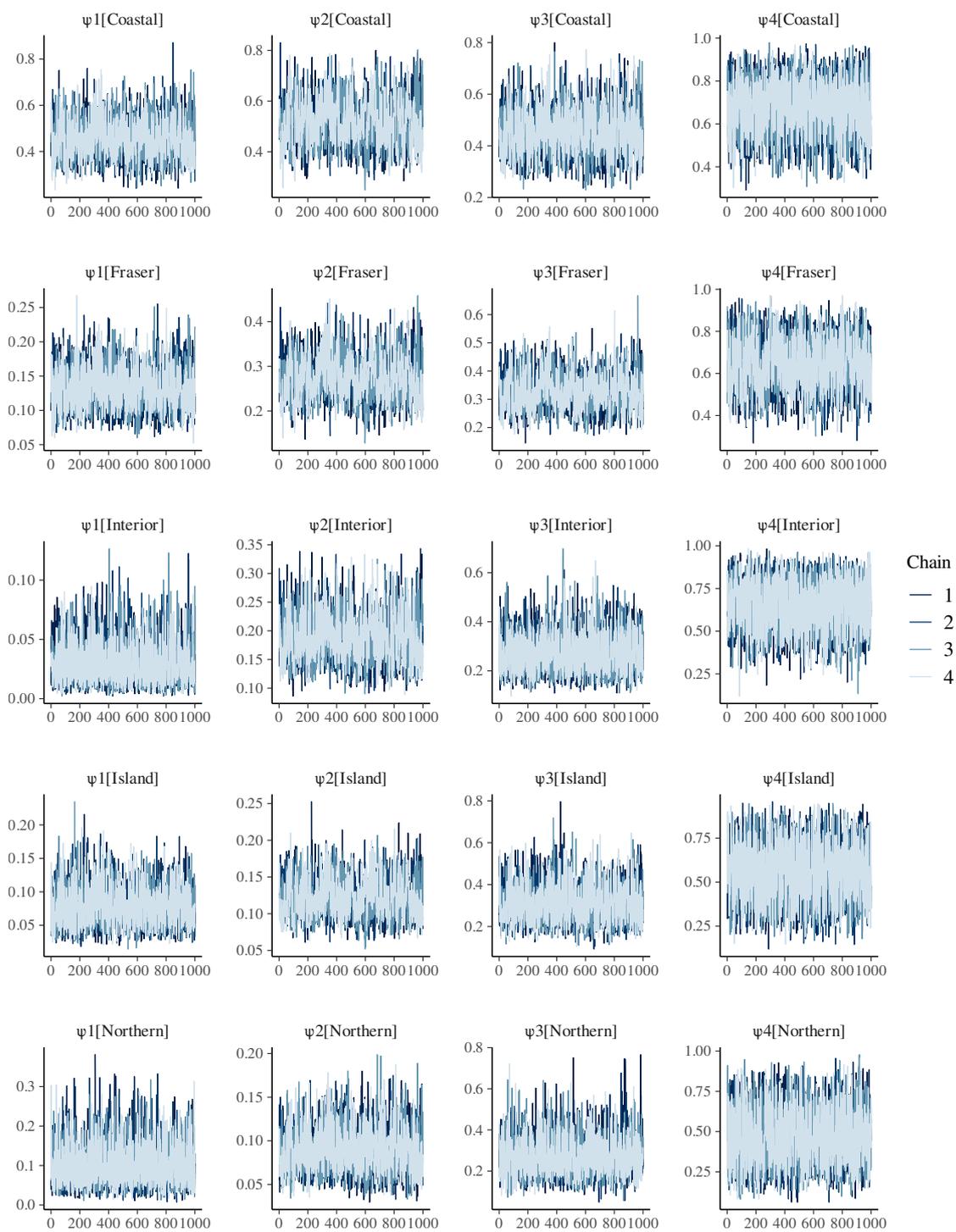}
    \end{subfigure}
    \caption{Trace plots of runs for the regional $\psi_1,\dots,\psi_4$.}
    \label{fig:traceplots-regional-a}
\end{figure}
\newpage
\begin{figure}[h!]
    \begin{subfigure}[b]{\textwidth}
        \centering
        \includegraphics[scale=0.7]{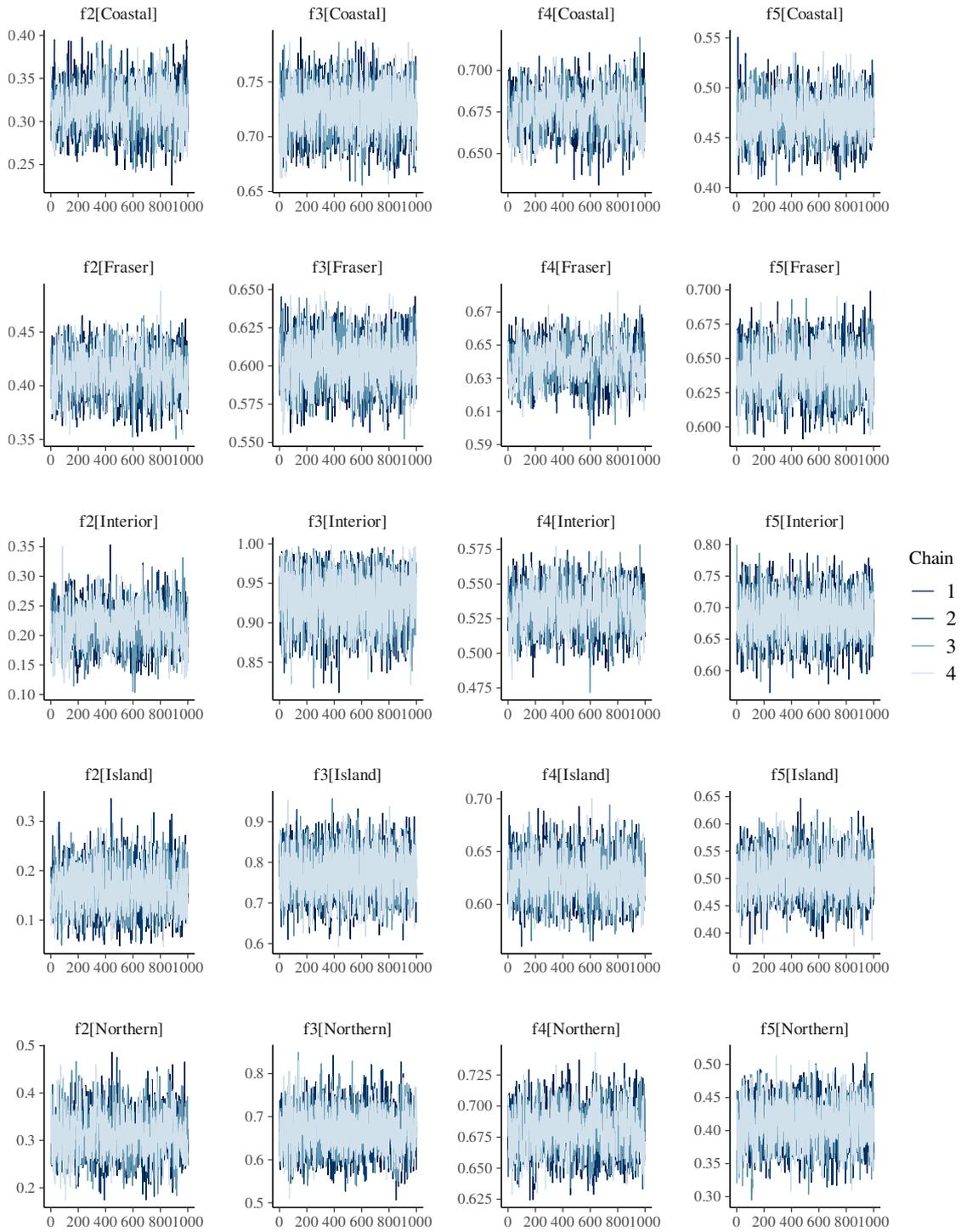}
    \end{subfigure}
    \caption{Trace plots of runs for the regional $f_2,\dots,f_5$.}
    \label{fig:traceplots-regional-b}
\end{figure}
\newpage
\begin{figure}[h!]
    \begin{subfigure}[b]{\textwidth}
        \centering
        \includegraphics[scale=0.7]{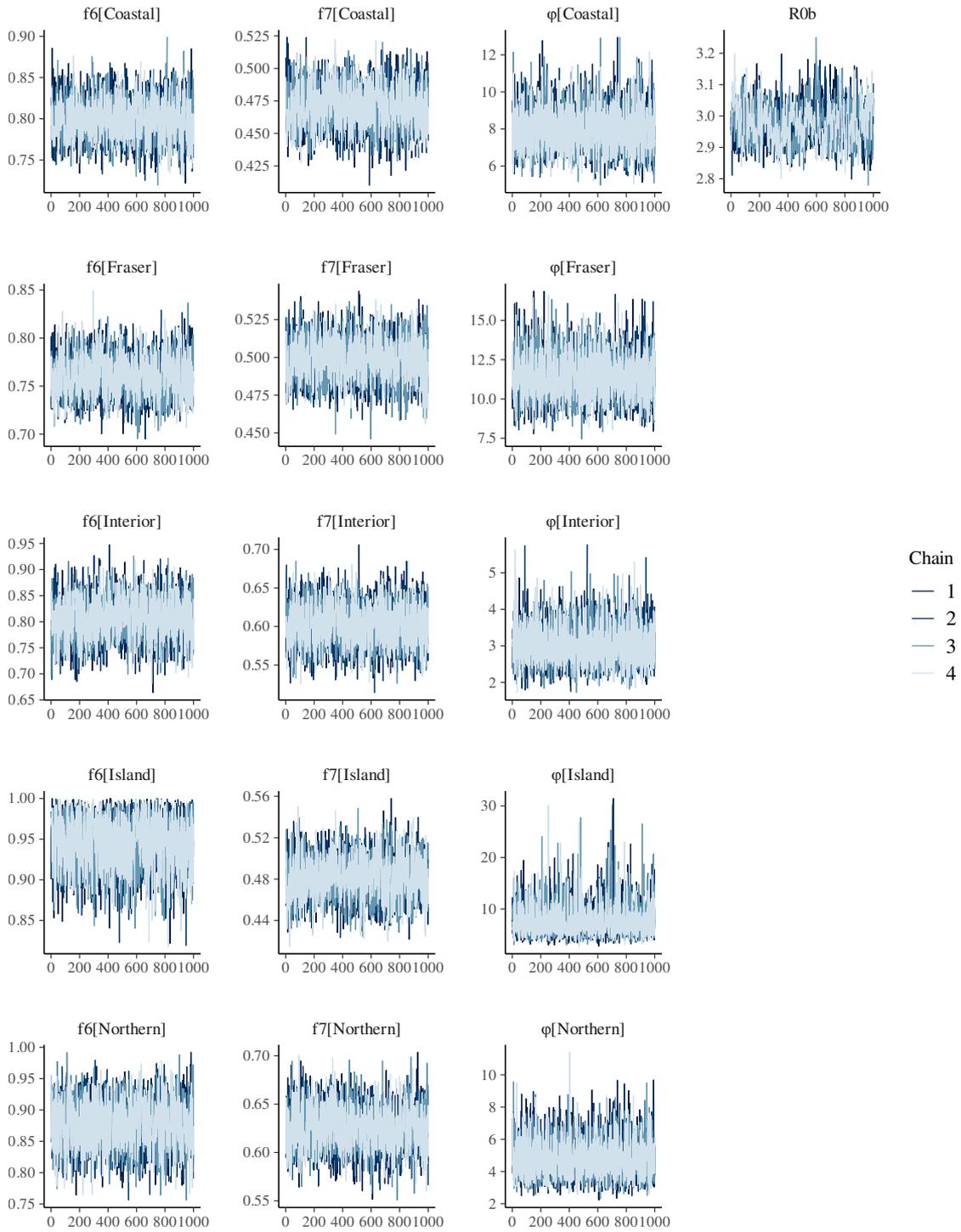}
    \end{subfigure}
    \caption{Trace plots of runs for the regional $f_6,f_7,\phi$ and hierarchical regional $R_{0b}$.}
    \label{fig:traceplots-regional-c}
\end{figure}
\newpage
\begin{figure}[h!]
    \begin{subfigure}[b]{\textwidth}
        \centering
        \includegraphics[scale=0.6]{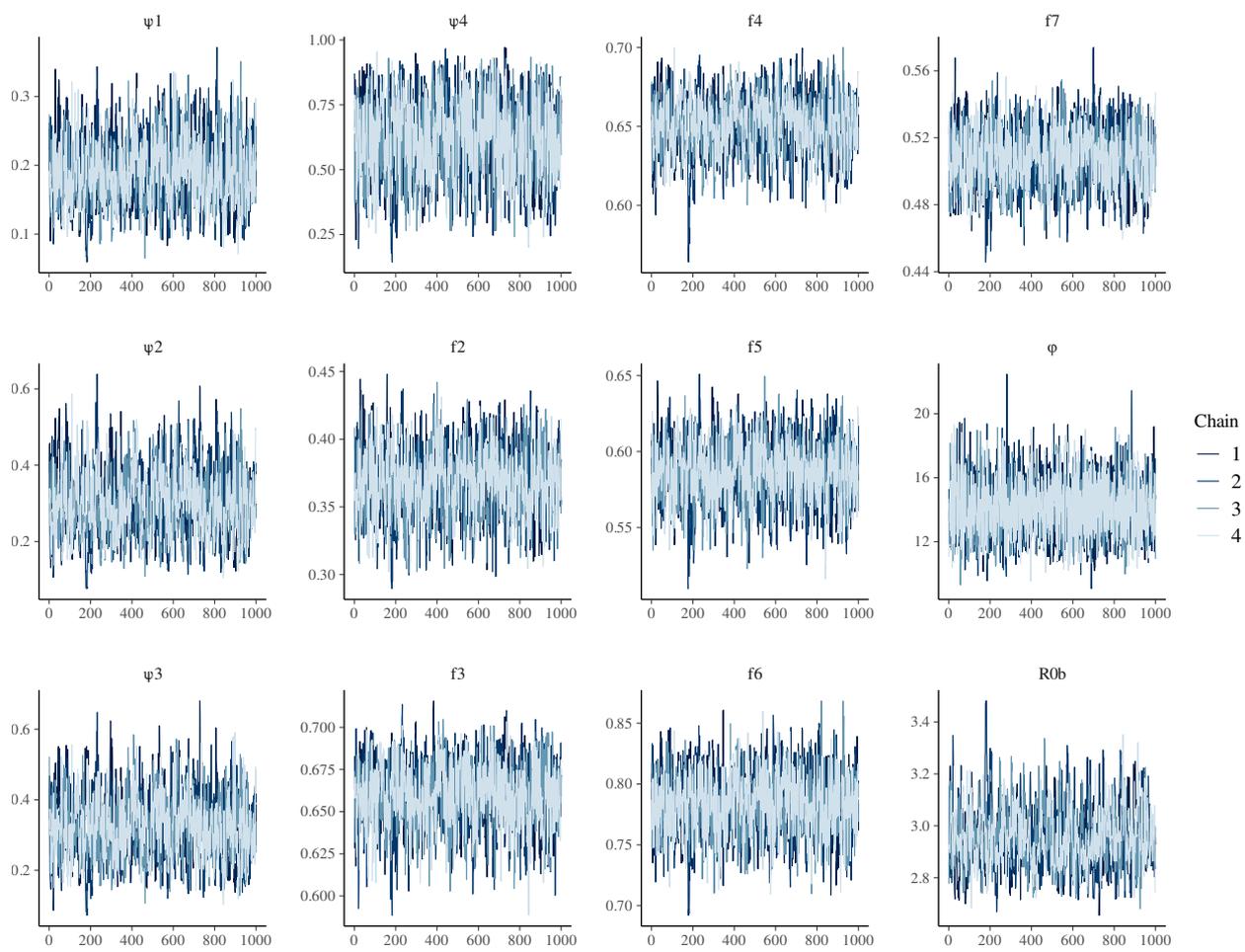}
    \end{subfigure}
    \caption{Trace plots of runs for the provincial-wide $\psi_1,\dots,\psi_4,f_2,\dots,f_7,\phi, R_{0b}$.}
    \label{fig:traceplots-BCwide}
\end{figure}
\end{appendices}

\end{document}